\documentclass[aps,prb,final,twocolumn,superscriptaddress,longbibliography]{revtex4-1}
\usepackage[caption=false]{subfig}

\usepackage[utf8]{inputenc}
\usepackage[T1]{fontenc}
\usepackage{graphicx}
\usepackage{grffile}
\usepackage{longtable}
\usepackage{wrapfig}
\usepackage{rotating}
\usepackage[normalem]{ulem}
\usepackage{amsmath}
\usepackage{textcomp}
\usepackage{amssymb}
\usepackage{capt-of}
\usepackage{hyperref}
\date{\today}
\title{}
\hypersetup{
 pdfauthor={Óscar Nájera},
 pdftitle={},
 pdfkeywords={},
 pdfsubject={},
 pdfcreator={Emacs 25.3.1 (Org mode 9.1.1)},
 pdflang={English}}
\begin{document}

\title{Multiple crossovers and coherent states in a Mott-Peierls insulator}

\author{O. Nájera}
\author{M. Civelli}
\affiliation{Laboratoire de Physique des Solides, CNRS-UMR8502, Université Paris-Sud, Orsay 91405, France}

\author{V. Dobrosavljevi\'c}
\affiliation{Department of Physics and National High Magnetic Field Laboratory, Florida State University, Tallahassee, FL 32306, USA}

\author{M. J. Rozenberg}
\affiliation{Laboratoire de Physique des Solides, CNRS-UMR8502, Université Paris-Sud, Orsay 91405, France}
\affiliation{Department of Physics, University of California San Diego, 9500 Gillman Dr. La Jolla 92093, USA }

\begin{abstract}
We consider the dimer Hubbard model within Dynamical Mean Field Theory to
study the interplay and competition between Mott and Peierls physics.  We
describe the various metal-insulator transition lines of the phase diagram
and the break down of the different solutions that occur along them. We
focus on the specific issue of the debated Mott-Peierls insulator crossover
and describe the systematic evolution of the electronic structure across
the phase diagram. We found that at low intra-dimer hopping the emerging
local magnetic moments can unbind above a characteristic singlet
temperature \(T^*\). Upon increasing the inter-dimer hopping subtle changes
occur in the electronic structure. Notably, we find Hubbard bands of a mix
character with coherent and incoherent excitations. We argue that this
state is relevant for VO\(_2\) and its signatures may be observed in
spectroscopic studies, and possibly through pump-probe experiments.
\end{abstract}

\maketitle

\section{Introduction}
\label{sec:org4c72b07}
Vanadates are fascinating materials that provide a fertile playground to
study non-trivial phase transitions driven by the electronic correlation.
Their structures follow the Magnéli series V\(_n\)O\(_{2n-1}\) and most of
them exhibit insulator to metal transitions (IMT) upon heating.
Significantly, these transitions involve a structural change but are
disconnected to magnetic ordering, with the exception of the most famous
members of the series, V\(_2\)O\(_3\) and VO\(_2\). In the former the
magnetic and structural transition occur at the same temperature, while in
the latter there is a structural transition but no magnetic ordering at any
temperature.

Vanadium dioxide is particularly appealing because its transition occurs
close to room temperature, around 340K, so it may enable practical
applications in novel electronic devices
\cite{inoue2008_taming_mott_transition_novel_mott_transistor,stoliar2017_leaky_integrate_and_fire_neuron}.
In fact, the study of this material is receiving a great deal of
attention. Therefore, a basic understanding of the nature of its electronic
state and its insulator to metal transition (IMT) is of great importance.

VO\(_2\) may be considered a realization of a Mott system, as it has only
one electron per vanadium atom which should lead to a partially filled band.
However, since it exhibits a non-magnetic insulating ground state, this
classification has often been challenged. In fact, in an influential early
paper, Goodenough
\cite{goodenough1960_direct_cation_interactions_several_oxides} argued that
the first order IMT occurs concomitant with a structural distortion of the
vanadium chains in the crystal, so the gap opening upon cooling should be
due to a Peierls lattice instability. Nevertheless, Density Functional
Theory (DFT) calculations failed to substantiate Goodenough's claim
\cite{wentzcovitch1994_vo,eyert2002_metal_insulator_transitions_vo2} so the
nature of the insulating state in VO\(_2\) remained a puzzle. The development
of methods that incorporate strong correlation effects in realistic lattice
calculations provided new light. Eventually, Biermann et al. showed using
Cluster Dynamical Mean Field Theory (CDMFT) with DFT that strong
correlations due to local Coulomb repulsion may lead to the opening of a
gap
\cite{biermann2005_dynamical_singlets_correlation_assisted_peierls_transition_invo2,tomczak2008_effective_bandstructure_insulating_phase_versus}.
Hence, the ground state was considered a ``Peierls insulator with dynamical
correlations''. This theoretical problem continued to attract theoretical
attention as the numerical techniques were further improved. Weber et
al. \cite{weber2012_vanadium_dioxide} argued that the mechanism driving the
insulator state was better characterized as a ``Peierls assisted orbitally
selective Mott transition''. They observed that mainly the a\(_{1g}\) orbital
drives the opening of the gap. A more recent
study where the oxygen atoms were explicitly included
\cite{brito2016_metal_insulator_transition_in_vo} enabled a more
comprehensive account of the various phases observed in VO\(_2\). However,
it also led to the reinterpretation of the transition as a ``Mott
transition in the presence of strong intersite exchange''. We should also
mention here the work by Eyert, whose calculations based on hybrid
functionals \cite{eyert2011_vo2_novel_view} may also open a gap in the
monoclinic phase.

While the technical improvements of the computational methods of realistic
correlated materials made significant steps forward in
our understanding of the electronic states, it is also true that their
technical complexity represents a challenge. For instance, despite multiple
studies dedicated to this material
\cite{biermann2005_dynamical_singlets_correlation_assisted_peierls_transition_invo2,belozerov2012_monoclinic_m_phase_vo,weber2012_vanadium_dioxide,brito2016_metal_insulator_transition_in_vo}
some basic issues remain unaccounted for, such as the
existence of a first order thermally driven insulator-metal transition just
above room temperature. Finite temperature studies are in principle beyond
the applicability of DFT methods although we should point out the recent
work of Pla\v{s}ienka et
al. \cite{plasienka2017_ab_initio_molecular_dynamics_study}. We investigated
the issue of a first order IMT in VO\(_2\) in a recent paper
\cite{najera2017_resolving_vo2_controversy} employing a model Hamiltonian
approach within CDMFT. Although the model is rather schematic, namely, a
lattice of correlated dimers, it was intended to qualitatively capture the
dimerization-delocalization competition in the monoclinic phase of
VO\(_2\). We showed that adopting model parameters relevant for VO\(_2\)
there is a thermally driven
first order Mott transition that occurs at a temperature range compatible
with the experimental observation. Moreover, we have also provided an
interpretation to the puzzling presence of a mid-infrared peak in the
optical conductivity of metallic nano-size islands observed during the
transition \cite{qazilbash2007_mott_transition_vo2_revealed_by} in terms of a
novel correlated ``monoclinic'' metal
\cite{najera2017_resolving_vo2_controversy}.  Model Hamiltonians that capture
the key ingredients are also attractive for
experimentalist as they provide useful insights within a more intuitive
yet non-trivial physical framework for strongly interacting systems.

In the present paper we continue with this strategy and make new inroads
into the investigation of the nature of the insulator and metallic states
realized in the Dimer Hubbard Model (DHM) within CDMFT. Specifically we
address the issue on the physical characterization of the insulator
state. A key
feature of our approach is to observe that the quantum impurity problem of
the DHM has the same form as the respective quantum impurity problem of the
DFT+CDMFT method on realistic lattices. In fact, we shall see that our
model solution does capture in a simpler context several features already
seen in DFT+CDMFT studies
\cite{tomczak2007_effective_band_structure_correlated_materials,tomczak2008_effective_bandstructure_insulating_phase_versus},
such as the presence of renormalized coherent bands in
the insulator.

The model Hamiltonian approach enables the systematic investigation of the
whole parameter space, which sheds light on the interplay of physical
mechanisms. We shall address the question of the physical crossover from a
pure Peierls insulator, which is a band insulator of a lattice of dimers
without correlations, to the pure (un-dimerized) correlated Mott
insulator. We shall see that the behavior of the system across the
crossover regime is non-trivial. One of our main results is that as the
system evolves from the Mott to the Peierls insulator there are at least
four different regimes, including an unprecedented state where the Hubbard
bands have an electronic structure with mixed coherent and incoherent
character. Interestingly, VO\(_2\) seems to be in this peculiar regime, which
might eventually be seen in spectroscopy experiments. On the other hand, we also
investigated the correlated ``monoclinic'' metallic state and show that it
can be understood in simple terms as a renormalized two-band heavy metal at
low frequencies. Interestingly, indications of a monoclinic metal in VO\(_2\)
have been reported in several experimental studies
\cite{kim2006_monoclinic_correlated_metal_phase_invo2as,arcangeletti2007_evidence_pressure_induced_metallization_process,qazilbash2011_nanoscale_imaging_electronic_structural_transitions_vanadium_dioxide,tao2012_decoupling_structural_electronic_phase_transitions_invo2,yoshida2014_ultrafast_photoinduced_transition_insulating_vo,morrison2014_photoinduced_metal_like_phase_monoclinic,laverock2014_direct_observation_decoupled_structural_electronic,wegkamp2014_instantaneous_band_gap_collapse_photoexcited}.

Consistent with the DFT+CDMFT studies, we find that VO\(_2\) should be
characterized as a realization of a Mott state in a dimer lattice
rather than a renormalized Peierls band insulator. However, in a larger
perspective, our systematic investigation of the model parameter space
should also shed light for the classification of a variety of
monoclinic transition metal oxide systems with the MO\(_2\) formula
\cite{hiroi2015_structural_instability_rutile_compounds} and the non-magnetic
insulator states of other vanadates.

This paper is organized as follows: In Section II we introduce the DHM and
the DMFT equations. We also describe a simple parametrization of a
renormalized two band model that will be useful for the discussion of our
results. In Section III we present the phase diagram and discuss the
various insulator-metal transitions of the model. In particular, we
describe the destruction of the correlated metal and the insulator in the
Mott regime. In Section IV we present the detailed study of the Mott to
Peierls crossover. We characterize the several distinct physical
regimes, including one with mixed coherent and incoherent features in the
Hubbard bands possibly relevant for VO\(_2\). Section V is dedicated to the
conclusions of our work.

\section{The dimer Hubbard model}
\label{sec:org34ebb63}

We focus on the dimer Hubbard Model, which is a basic and natural
extension of the single band Hubbard Model.

The DHM reads,

\begin{equation}
\label{eq:dimer-hubbard-model}
\begin{aligned}
H=& \left[-t \sum_{\langle i, j\rangle \alpha\sigma}
c^\dagger_{i\alpha\sigma} c_{j\alpha \sigma} +\ t_{\perp}
\sum_{i\sigma} c^\dagger_{i1\sigma} c_{i2\sigma} + H.c. \right] \\
& + U \sum_{i\alpha} n_{i\alpha\uparrow} n_{i\alpha\downarrow}
\end{aligned}
\end{equation}
where \(\langle i, j\rangle\) denotes nearest-neighbor lattice cells,
\(\alpha=\{1,2\}\) denote the dimer orbitals within a given cell, \(\sigma\) is the spin,
\(t\) is the lattice (i.e. inter-dimer) hopping and (\(t_\perp\)) is the
intra-dimer hopping. The parameter \(U\) is the on-site Coulomb repulsion.

\begin{figure}[htbp]
\centering
\includegraphics[width=\linewidth]{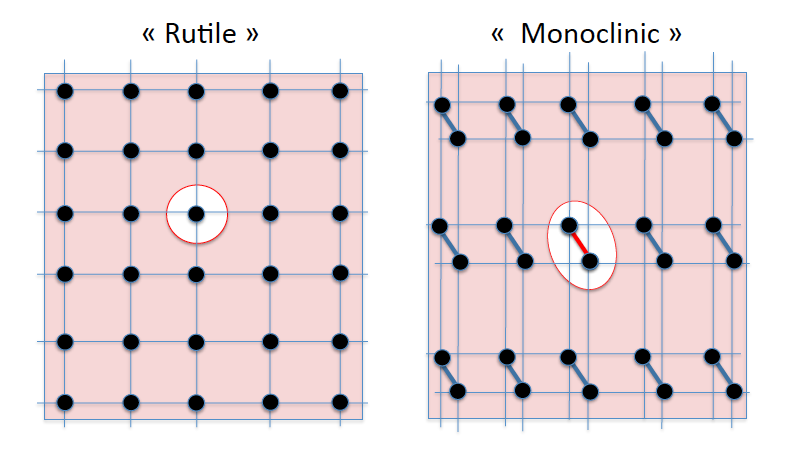}
\caption{Schematic representation of the higher symmetry ``rutile'' lattice with one atom per unit cell and the low symmetry ``monoclinic'' lattice, which is a lattice of dimers. In shaded red, we represent the quantum impurity effective environment determined through the CDMFT equations. Notice that our model is defined on a semi-circular non-interacting density of states, that may be realized in a Bethe lattice. Nevertheless, here we depict a square lattice just for the sake of simplicity. \label{fig0}}
\end{figure}

The non-interacting limit of the DHM has two bands which are locally
hybridized at every lattice cell site. This leads to the parallel splitting of
the two bands by \(2t_\perp\). When this splitting is large enough, the
system has a continuous metal insulator transition. We associate this to a
Peierls-like mechanism, as it is driven by the increase of the intra-dimer
hopping amplitude \(t_\perp\). It can be ascribed to a schematic
representation of the monoclinic distortion in the real material that
creates dimers in the unit cell (see Fig.~\ref{fig0}). Another way to see
this IMT is by starting from the \(t \to 0\) limit, where the local dimers
that form a bonding (B) and an anti-bonding (AB) molecular orbital at every
1-2 link. Switching on the inter-dimer hopping, these orbitals lead to an
insulating state with two split flat bands. At a large enough (lattice-geometry
dependent) value of \(t\), the B and AB bands start to overlap and realize
a transition into a metal.

Before proceeding, we should avoid any confusion here by noting that the
model \eqref{eq:dimer-hubbard-model} is fully defined only after its
lattice is specified. For instance, if dimers are arranged as a one
dimensional system, the model is a ``ladder'' (the dimer rungs are
perpendicular to the direction of the lattice). In two dimensions, as in
the schematic Fig.~\ref{fig0}, one would get a ``bi-layer'' model, where the
dimer rungs connect at every site two parallel 2D layers. Those systems
have qualitatively different behaviors from the one that concerns us here,
namely, the physics of the three-dimensional systems such as VO\(_2\). In 3D systems
that have strong local interactions, one may expect the Dynamical Mean
Field Theory (DMFT) to be a reasonable approximation \cite{georges1996_dynamical_mean_field_theory_strongly}. In fact, DFT+DMFT
methods are implicitly based on such an assumption. The DMFT approach to
the DHM with a dimer unit cell is, strictly speaking, a cluster-DMFT
calculation, possibly the simplest instance of CDMFT. In DMFT and CDMFT it
is mathematically convenient and thus customary to adopt a non-interacting
semi-circular density of states (DOS), which is realized in a Bethe
lattice \cite{georges1996_dynamical_mean_field_theory_strongly}. In addition,
such a DOS qualitatively resembles that of a three-dimensional lattice
system. In fact, it has a finite bandwidth given by \(4t\) defined in equation
\eqref{eq:dimer-hubbard-model}. As in previous works, we adopt as unit of
energy the half bandwidth \(D=2t=1\). We should emphasize here that the
physics of models treated within DMFT in general do not depend on specific
geometry of the lattice, but on the nature of the quantum impurity
problem. As we mentioned already, the realistic VO\(_2\) lattice within
DFT+CDMFT has the same type of quantum dimer impurity as in the present
case. Our model has the additional simplification of considering one
orbital at each site instead of the 3 in the realistic case. We may see in
Fig.~\ref{fig0} schematic ``rutile'' and ``monoclinic'' lattices. From our
previous discussion, the key feature is that in the former case there is a
single site in the unit cell, while in the latter case there is a
dimer. The ``rutile'' lattice can be qualitatively associated to a conventional single-band
Hubbard model \cite{georges1996_dynamical_mean_field_theory_strongly}. While
the ``monoclinic'' one can be thought of as two copies of Hubbard models
coupled at every unit cell site by the intra-dimer hopping \(t_\perp\). In
the limit of \(t_\perp \to 0\) the two copies become independent and one
recovers the conventional single site Hubbard model physics
\cite{georges1996_dynamical_mean_field_theory_strongly,moeller1999_rkky_interactions_mott_transition,najera2017_resolving_vo2_controversy}.
As an additional remark, here we should say that both the single site
Hubbard and the Dimer Hubbard models at half filling have
anti-ferromagnetic ground-states, which are favored in bipartite
lattices. Nevertheless, the study of the MITs within the PM (meta-stable)
states is important in its own right as it has been very
useful to reveal the physical competition between different correlated
states \cite{georges1996_dynamical_mean_field_theory_strongly}.

\subsection{CDMFT equations and the bonding anti-bonding basis}
\label{sec:org22c96f8}

The solution of the CDMFT equations are obtained in terms of the
one-particle propagators \(G_{\alpha,\beta}\) (with \(\alpha, \beta=
1,2\)), which are subject to the self-consistent condition

\begin{equation}
 {\bf G}(\omega)= \int d\varepsilon\rho(\varepsilon) [ (\omega -
\varepsilon) \mathbb{I} - t_\perp \sigma_x -\mathbf{\Sigma}(\omega)]^{-1}
\label{eq:MFE}
\end{equation}

where \(\sigma_x\) is the \(x\) Pauli matrix and \(\Sigma_{\alpha,\beta}\)
is the local self-energy. The calculation of the local self-energy requires
the solution of the so-called local quantum impurity problem, which are
often generalizations of the Kondo problem
\cite{georges1996_dynamical_mean_field_theory_strongly}. To solve such a
quantum impurity problem requires the numerical implementation of a
many-body approach. Following previous work
\cite{najera2017_resolving_vo2_controversy}, here make extensive use of the
Iterative Perturbation Theory (IPT)
\cite{moeller1999_rkky_interactions_mott_transition} that allows fast and
precise calculations in the whole phase diagram and at all \(T\) and
\(T=0\). Notably, IPT becomes \emph{exact} in the limits of \(T=0\) and \(t \to
0\) or \(U \to 0\), \emph{for all} \(t_\perp\) (see Appendix B) and its numerical precision allows
for reliable analytic continuation of the data to the real axis.  We also
benchmarked our IPT calculations with results from other numerical methods
such as the continuous time quantum Monte Carlo (CT-QMC)
\cite{werner2006_hybridization_expansion_impurity_solver,seth2016_triqs} and
the exact diagonalization (ED)
\cite{georges1996_dynamical_mean_field_theory_strongly}. The former is a
finite \(T\) calculation and the latter is a \(T=0\) one. These methods are
both exact up to systematic errors, but are numerically very expensive and
have the serious drawback of requiring, in the case of CT-QMC, the analytic
continuation of the (noisy) imaginary axis results.

In the present case of half-filling (i.e., one particle per site),
particle-hole symmetry holds.  We further assume translational invariance
and search for a paramagnetic solution. We have then:
\(G_{11}(\omega)=G_{22}(\omega)\), and \(G_{12}(\omega)=G_{21}(\omega)\).
The self-energies have similar properties.  In order to simplify the
discussion of the evolution of the electronic structure, it is convenient
to consider the B/AB representation that renders the Green's Functions and
Self-energies \emph{diagonal}.

\begin{subequations}
\label{eq:diagonal_bands}
\begin{align}
\label{eq:diagonal_gf}
G_{B/AB} &= G_{11}\mp G_{12}\\
\label{eq:diagonal_sigma}
\Sigma_{B/AB} &= \Sigma_{11}\mp \Sigma_{12}
\end{align}
\end{subequations}

In the B/AB basis the electronic structure of the non-interacting problem
is particularly simple. The single particle energies \(E_k^{B/AB}\) form
two parallel bands of bandwidth \(2D\) and are split by \(2t_\perp\), i.e.,
\(E_k^{B/AB} = \mp t_\perp + \epsilon_k\), where \(\epsilon_k\) is the
single particle energy of the bands for \(t_\perp = 0\). In the present
case, since we employ a semi-circular DOS (realized in a Bethe lattice in
infinite dimensions) the lattice single particle energies \(\epsilon_k\)
drop their \(k\)-dependence
\cite{georges1996_dynamical_mean_field_theory_strongly} and then are simply labeled by
\(\epsilon \in [-D,D]\)
\cite{georges1996_dynamical_mean_field_theory_strongly}. Thus, at finite
\(t_\perp\) and \(U=0\), the model DOS is composed of two semicircles split
by \(2t_\perp\), \(\rho_{B/AB}(\omega) = \frac{2}{\pi {D}^2}
\sqrt{{D}^2-(\omega\pm {t}_{\perp})^2}\).

\subsection{The renormalized two band model}
\label{sec:orgbfa56e3}

In order to better analyze our results in the subsequent sections, it is
convenient to introduce here a simple low-energy parametrization of the
two-band system.  We may think of this as a non-interacting renormalized
two-band model (R2B).  In a normal (i.e. Fermi liquid) metal the
self-energy is well-behaved at low frequencies\footnote{The derivation of the R2B formulas and a study of the behavior of
the Self-energy in the low frequency regime is presented in appendix
\ref{r2b_approx}}. In the present
case we expand around \(\omega =0\) the self-energies of the mean-field
equations \eqref{eq:MFE}, and introduce a quasiparticle residue \(Z\):

\begin{equation}
\label{eq:Z}
Z^{-1} = 1-\frac{\partial\Re \Sigma_{B/AB} (\omega)}{\partial \omega} \bigg |_{0} =
1-\frac{\partial\Re \Sigma_{11} (\omega)}{\partial \omega} \bigg |_{0}
\end{equation}

and a renormalized intra-dimer hopping \(t_\perp\) as:

\begin{equation}
\label{eq:tpeff}
\tilde{t}_\perp = t_\perp \mp \Re \Sigma_{B/AB}(\omega) |_{0} = t_\perp + \Re \Sigma_{12}(\omega) |_{0}
\end{equation}

The DOS of the R2B model then reads,

\begin{equation}
\label{eq:low_energy_dos}
\rho^{R2B}_{B/AB}(\omega) \sim \frac{2}{\pi D^2}\,
\sqrt{D^2 - \big( \frac{\omega}{Z} \pm \tilde{t}_{\perp}\big)^2}
\end{equation}
that corresponds to two heavy effective bands with dispersion
\(E_\epsilon^{B/AB} = \mp Z\tilde{t}_\perp + Z\epsilon\), where the
effective mass renormalization is \(m^*/m = 1/Z\). The overlap between
the two band-edges is given by \(2\eta\), where

\begin{equation}
\eta = ZD - Z\tilde{t}_\perp .
\label{eq:transition_orderparam}
\end{equation}

\begin{figure}[htbp]
\centering
\includegraphics[width=\linewidth]{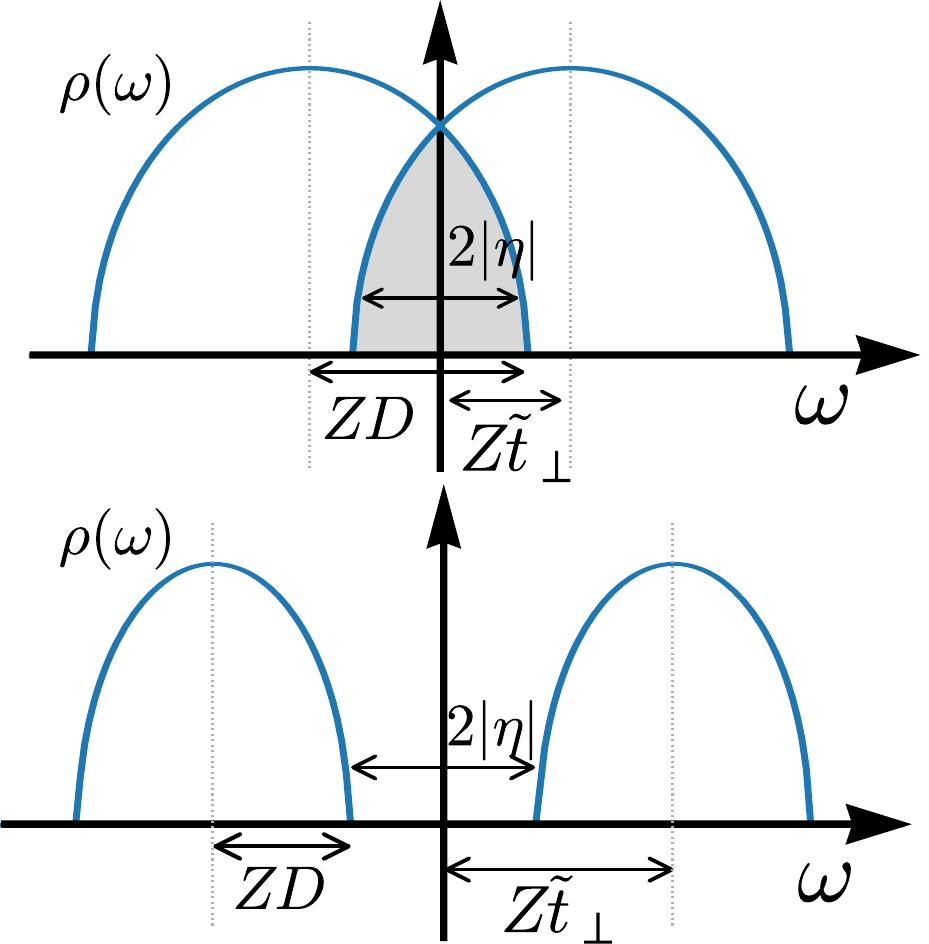}
\caption{Schematic representation of the two bands in the R2B model. This model is a simple renormalization of the non-interacting case where \(Z=1\) and the intra-dimer hopping is \(\tilde{t}_\perp=t_\perp\). \(ZD\) denotes the renormalization of bandwidth and \(2\eta\) the overlap of the two bands. \label{fig:R2B}}
\end{figure}

Thus, for \(\eta>0\) we have a metal state, and for
\(\eta<0\)(i.e. splitting dominates) the DOS
opens a gap (see Fig. \ref{fig:R2B}). We will use this quantity to describe
the metal to insulator transition, in the next section.

The renormalized two-band (R2B) model may describe both metallic and
insulating states, so long the \(\Sigma\) remains well-behaved (i.e. linear) according to
the parametrization. As it turns out the description will be valid
at low frequencies throughout the metallic phase, which is a Fermi
liquid. In the insulator, we shall see that it is a good approximation only
within the Peierls limit, where the interaction and thus the \(\Sigma\) are
small. In the Mott regime this
parametrization is not appropriate.

\section{Phase diagram and insulator-metal transitions}
\label{sec:orgcd05f74}

\begin{figure}[htbp]
\centering
\includegraphics[width=\linewidth]{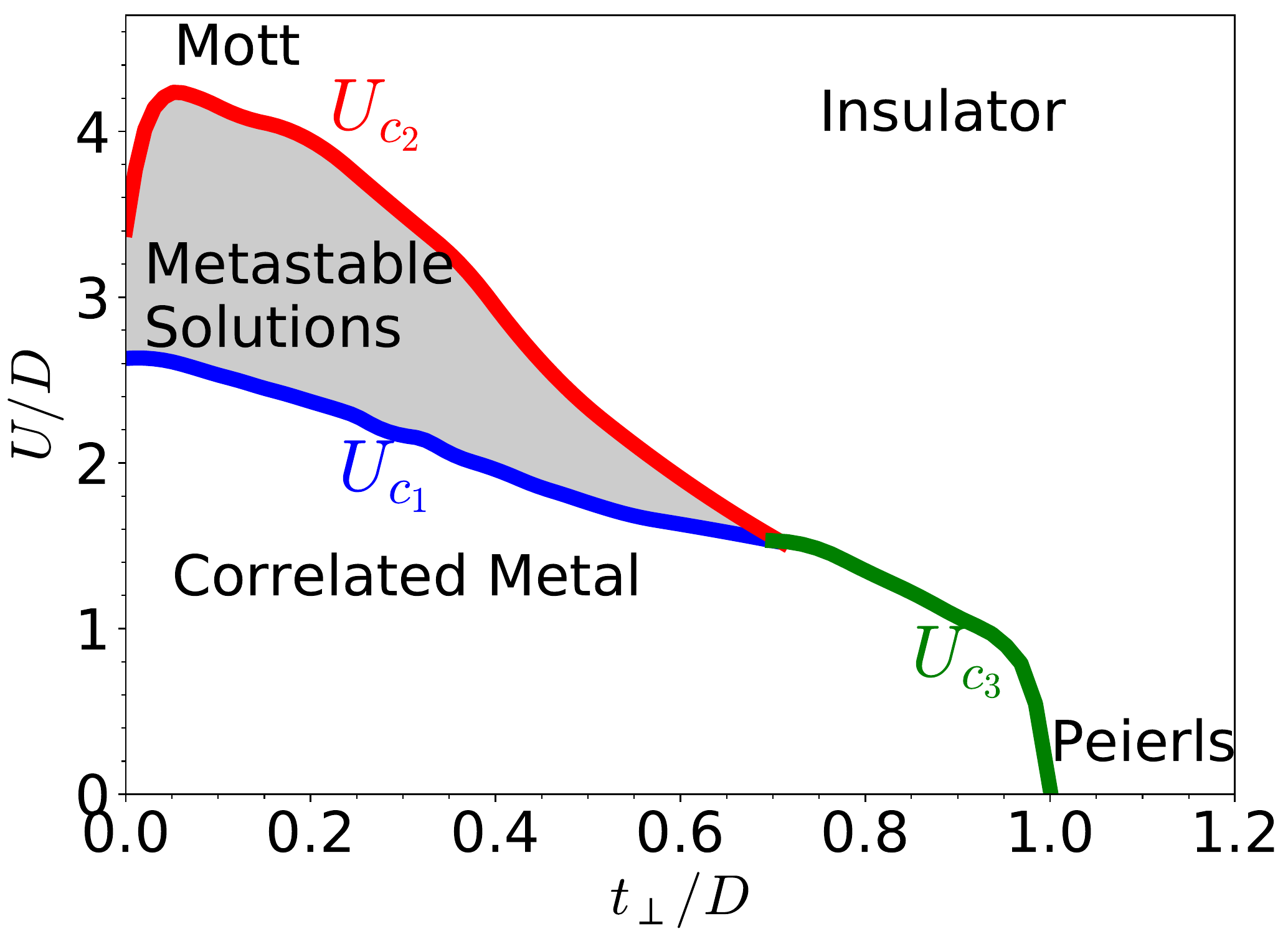}
\caption{The Metal-Insulator transition takes place along the \(U_{c_1}\), \(U_{c_2}\) and \(U_{c_3}\) lines (blue, red and green respectively). The \(U_{c_2}\) line corresponds to a spinodal line where the metallic solution discontinuously disappears upon increasing \(U\). Along the \(U_{c_3}\) line there is a continuous 2nd order metal-insulator transition. The \(U_{c_1}\) line marks the spinodal where the Mott insulator vanishes discontinuously upon decreasing \(U\). The Mott insulator is continuously connected to the Peierls insulator, however different crossover behaviors can be identified. \label{fig:phasediag}}
\end{figure}

In Fig.~\ref{fig:phasediag} we show the \(U-t_\perp\) phase diagram at \(T=0\),
which is well-known from previous studies
\cite{moeller1999_rkky_interactions_mott_transition,hafermann2009_metal_insulator_transition_by_suppression_spin_fluctuations,fuhrmann2006_from_mott_insulator_to_band_insulator,najera2017_resolving_vo2_controversy}.
We recall now its main features. There is a metallic phase for \(t_\perp <
1\), and an insulator phase at high enough \(U\). The IMT changes character
depending on the value of \(t_\perp\). At values higher than \(t_\perp \approx
0.7\) the transition is continuous (2nd order) along a line \(U_{c_3}\)
indicated in green in Fig.~\ref{fig:phasediag}. At smaller values of \(t_\perp\)
there are two lines \(U_{c_1}\) and \(U_{c_2}\) respectively indicated in blue
and red in the figure. These are two spinodal lines of the mean field
theory self-consistent solution. The metal state is destabilized for
\(U>U_{c_2}\) whereas the insulator state
is destabilized along \(U<U_{c_1}\). Thus, in between the two spinodal lines
there are two different solutions of the CDMFT equations, one metallic and
one insulating. At finite temperature this coexistence region shrinks,
until it disappears at a critical temperature. At
higher \(T\) there is a crossover behavior and bad metal states
\cite{najera2017_resolving_vo2_controversy}.  This phase diagram is obtained
with IPT but we have validated all its main features by extensive CT-QMC
calculations \cite{najera2017_resolving_vo2_controversy}.

The description of how these transitions take place in this basic model of
strongly correlated systems has (rather surprisingly) not been investigated
in detail. As we discussed before, this type of impurity model is at the core of calculations
of realistic material with a dimer in the unit cell. We shall therefore
describe the transitions in detail in the next subsections.

\subsection{Metal to insulator transitions across \(U_{c_2}\) and \(U_{c_3}\)}
\label{sec:org1c3ef86}

\begin{figure}[htbp]
\centering
\includegraphics[width=\linewidth]{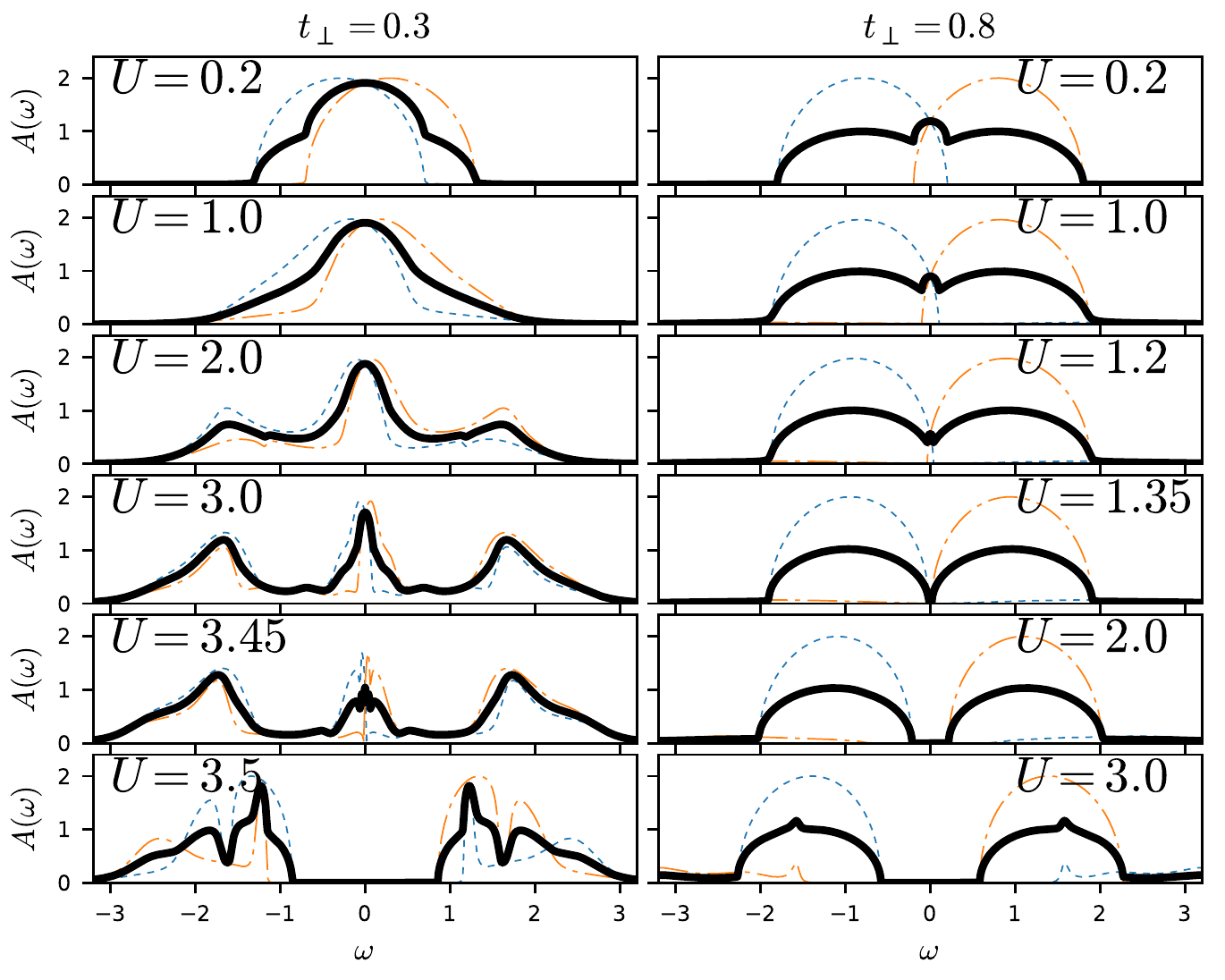}
\caption{DOS for increasing values of \(U\) crossing the red \(U_{c_1}\) and the green \(U_{c_3}\) boundary lines (left and right panels, respectively) of the phase diagram of Fig.~\ref{fig:phasediag}. Black lines represent the local(\(-\Im m G_{11}\)) Spectral function, which is the average of the bonding (dashed blue) and anti-bonding (dot dashed orange) bands, calculated at \(T=0\) with IPT. \label{fig:dimer_transition_spectra}}
\end{figure}

The metal to insulator transition by increasing \(U\) dramatically changes
its character as a function of \(t_\perp\). In
Fig.~\ref{fig:dimer_transition_spectra} we illustrate this by showing the
evolution of the frequency-dependent DOS with increasing \(U\), for two
representative values \(t_\perp=0.3\) and 0.8, that respectively cross the
\(U_{c_2}\) and the \(U_{c_3}\) lines.

The behavior of the DOS at higher \(t_\perp= 0.8\) displays a rather simple
evolution. As shown by the data in the B/AB basis (dotted blue and
dot-dashed orange) we observe that the total DOS (thick black) is the
average of the two semicircular contributions of the B and AB bands. The
gap opens continuously and the effective masses or, equivalently, the bandwidth of the two
bands remains essentially un-renormalized. The insulator at \(U>U_{c_3}\) is
clearly a band insulator state. As we mentioned before, we identify this
state with the Peierls insulator, since it is realized at large \(t_\perp\)
and relatively low \(U\). We may also note that at the highest values of the
interaction \(U\) this simple description begins to fail as the B and AB DOS
begin to develop a second contribution to the spectral weight for \(\omega\)
> 0 and < 0, respectively. We shall discuss this feature in more detail latter on.

In contrast, the transition at lower values of \(t_\perp= 0.3\) is
significantly different. The evolution of the DOS is more complex and has
various contributions. We can immediately observe the strongly correlated (Mott-Hubbard) character by
noticing a characteristic 3-peak structure at intermediate values of the
interaction.  The central peak, which gives the metallic character becomes
narrower as it losses spectral weight that is transferred to build the high
energy Hubbard bands at energy of order \(\pm U/2\). Interestingly and
contrary to the single band Hubbard model case, there is no pinning
condition for the central quasiparticle peak
\cite{moeller1999_rkky_interactions_mott_transition} and this quasiparticle
develops a non-trivial structure in the DOS at low frequencies as the
critical value \(U_{c_2}\) is approached. We shall come back to this point
also later on.  Unlike the higher \(t_\perp\) case, the decomposition of the
DOS in the B/AB contributions do not seem to provide any simpler picture of
the evolution.

\begin{figure}[htbp]
\centering
\includegraphics[width=\linewidth]{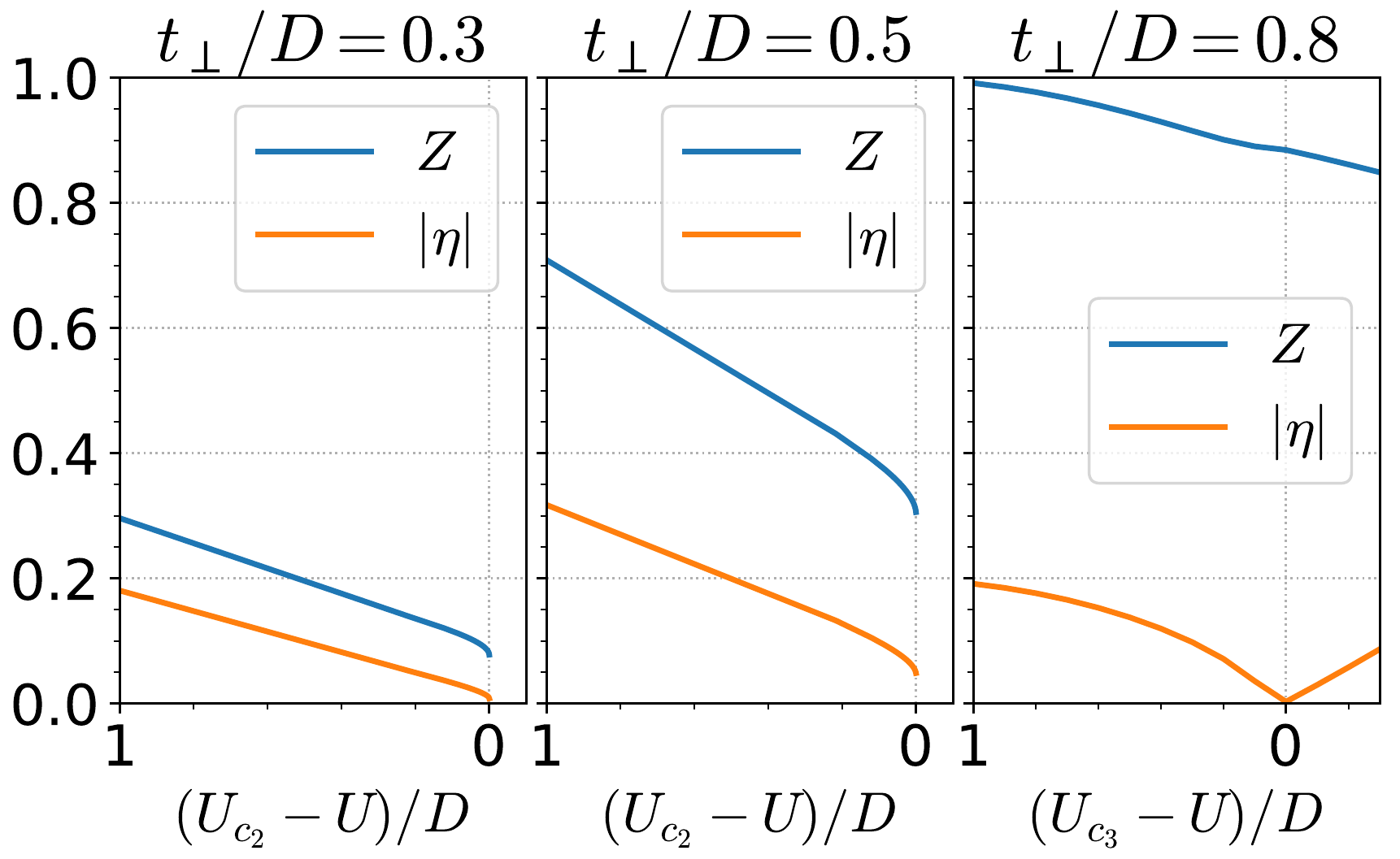}
\caption{\(Z\) and \(| \eta |\) as a function of \(U\) for various values of dimerization \(t_\perp\). The left and central panels correspond to the behavior as the \(U_{c_2}(t_\perp)\) red spinodal line is approached (cf. Fig.\ref{fig:phasediag}), where the metal-insulator transition is discontinuous. The right panel shows the behavior when the green 2\(^{\text{nd}}\) order line is crossed (cf. Fig.\ref{fig:phasediag}). The R2B parametrization works on either side of the transition. Note that \(\eta\) is negative on the insulating side in the last panel. \label{fig:transition_orderparam}}
\end{figure}

We can gain further insight on this transition by tracking the behavior of
the self-energy through the two parameters that we defined above for the
R2B model, \(Z\) and \(\eta\), that we show in
Fig.~\ref{fig:transition_orderparam}.

Consistent with our previous discussion, in the \(t_\perp= 0.8\) case we see
the parameter \(|\eta|\) continuously and linearly going to zero at both
sides of the transition (\(\eta > 0\) in the metal). The parameter \(Z\)
remains close to one, which indicates an almost negligible mass
enhancement. As one lowers the value of \(t_\perp\) to 0.5, we observe that
the \(Z\) parameter begins to experience a bigger renormalization. Further
down, for \(t_\perp=0.3\) the mass renormalization (\(\propto 1/Z\)) is very
large. Thus the metal state is strongly correlated with heavy
quasiparticles. We can think of such a state at low energies as resulting
from two ``Kondo'' states at each one of the atomic impurity sites. Each
one of the sites is independently screened by conduction electrons, and
also by each other. This leads to renormalized bonding and anti-bonding
heavy bands as was already discussed in our previous paper
\cite{najera2017_resolving_vo2_controversy}. Here we shall be concerned with
the question of how this heavy metallic state breaks down as \(U\) is
increased.

A heavy metal with a (single) band and divergent mass is a hallmark of the
metal-insulator transition in the single band Hubbard model within DMFT
\cite{georges1996_dynamical_mean_field_theory_strongly}. That transition is a
realization of the well-known Brinkman-Rice scenario, where the effective
mass diverges at the metal-insulator transition. In the present dimer model
we shall see that, despite a large renormalization of the effective mass,
it does not diverge and the transition is qualitatively different.

\begin{figure}[htbp]
\centering
\includegraphics[width=\linewidth]{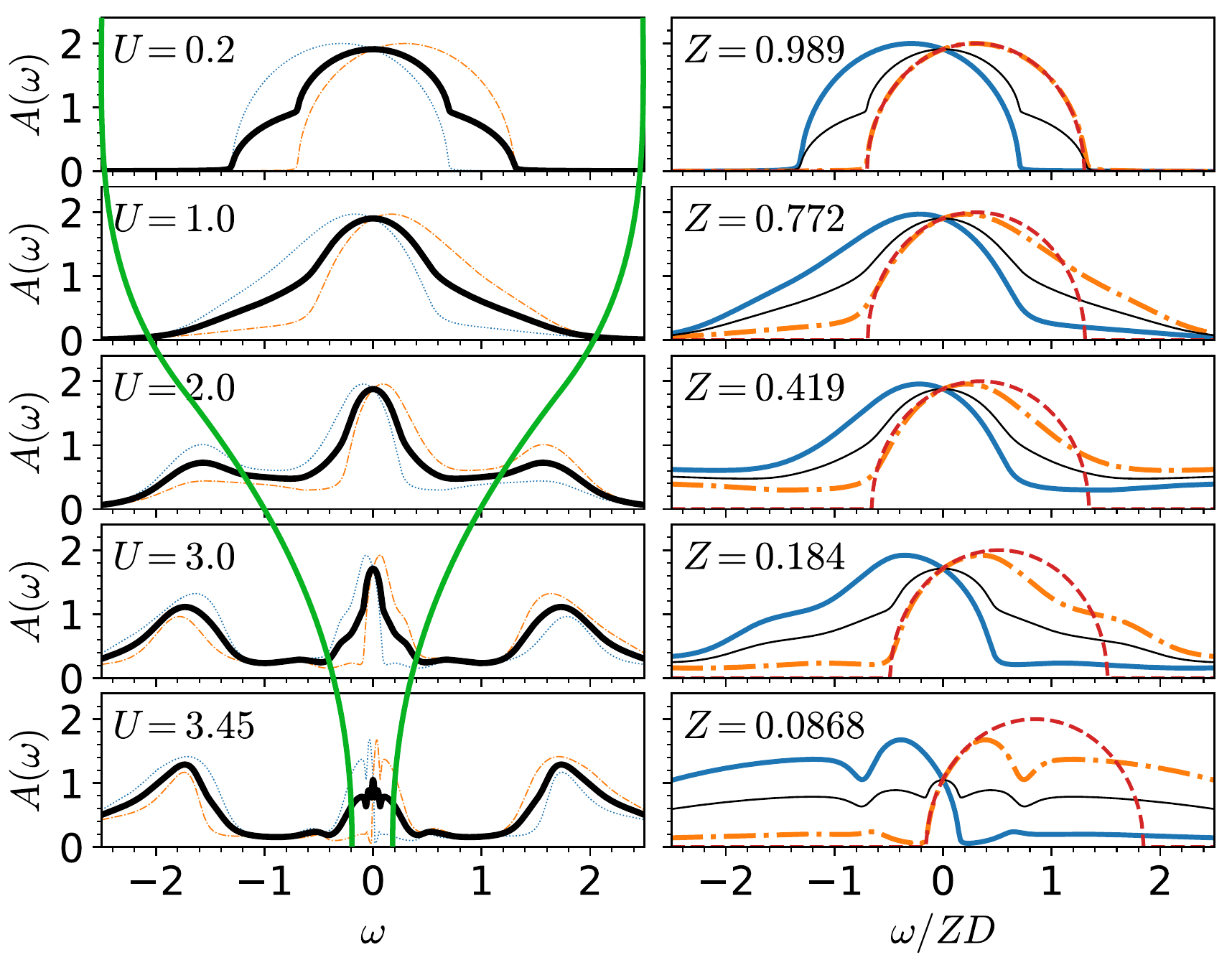}
\caption{Low frequency scaling of the metallic DOS from Fig.\ref{fig:dimer_transition_spectra} for increasing \(U\) at fixed \(t_\perp =0.3\). Right panels show a zoom into the corresponding low frequency region that is indicated with green lines in the left panels. The local spectral function (black lines) is decomposed in the Bonding Spectral function (Blue line) and Anti-bonding Spectral function(orange dot dashed line). The dashed red line is the renormalized parametrization of the low energy quasiparticle from equation \eqref{eq:low_energy_dos}. The superposition of the two bands decreases as the system approaches the critical value \(U_{c_2}\approx 3.47\). The size of the superposition is the vanishing scale \(\eta\), which in this scaled plot is \(\eta/ZD = 1 - \tilde{t}_\perp/D\). \label{fig:transition_spectra_scaling}}
\end{figure}

Motivated by the previous discussion and by Eqs.\ref{eq:tpeff} and
\ref{eq:low_energy_dos} of the renormalized two band model, we close-up on
the low frequency peak of the spectra of
Fig.~\ref{fig:dimer_transition_spectra} and we replot the DOS in
Fig.~\ref{fig:transition_spectra_scaling} as a function of the rescaled
frequency \(\omega/ ZD\). We see that a clear picture emerges, where the
central peak can be understood as two bands whose splitting is renormalized
down but whose width is also renormalized down.  The R2B model
parametrization (in dashed red line in the figure) provides a good
representation of the lowest frequency part of the spectra, made by the low energy
edges of the B and AB peaks. Otherwise, is not very accurate.

Unlike in the Brinkman-Rice scenario, where \(Z \to 0\), here the transition
occurs at a finite effective mass. The opening of the gap results from the
combined effect of the renormalization of the bandwidth and of the
splitting. Both decrease as \(U \to U_{c_2}(t_\perp)\), but the quantity that
becomes zero is not \(Z\) but the renormalized B/AB band overlap \(\eta =
ZD - Z\tilde{t}_\perp\). This means that the low-energy B and AB
contributions to the quasi-particle peak separate. This behavior is similar
to the MIT reported in a correlated two orbital model
\cite{mazza2016_field_driven_mott_gap_collapse}.  Despite the lack of a mass
divergence, the transition does share a similarity with the MIT in the
single band case, namely, that as the DOS(\(\omega=0\)) becomes zero the
Kondo effect can no longer be sustained and the impurities lose their
respective Kondo screening clouds. In the single band (single-site) case, in the Mott insulator state one is
left with almost free local moments. However, in the present situation a strong
RKKY-like magnetic interaction between the two sites takes over and one has
intra-dimer magnetic screening.  This dramatic enhancement of the intra-dimer
magnetic interaction translates into the sharp increase of the intra-dimer
effective hopping \(\tilde{t}_\perp = t_\perp + \Re\Sigma_{12}(0)\), which
drives the opening of the gap (see Fig.~\ref{fig:dimer_transition_spectra})
\cite{najera2017_resolving_vo2_controversy}. Interestingly this sharp
increase of the \(\Re\Sigma_{12}(0)\) was also observed in DFT+CDMFT
calculations in VO\(_2\) by Brito et
al. \cite{brito2016_metal_insulator_transition_in_vo}. They related this
effect to the gap-opening by the B/AB band splitting of the a\(_{1g}\)
orbital in VO\(_2\). It is interesting to see that our present model
Hamiltonian does capture the same basic physical mechanism, albeit in a
simplified scenario that makes its physical interpretation transparent. Thus, we see
how the renormalization of the intra-dimer hopping that drives the MIT
originates in the loss of lattice Kondo screening and the concomitant boost
of the local intra-dimer
magnetic interaction. These competing mechanisms are well known in strongly
correlated systems tracing back to Doniach's Kondo lattice
\cite{doniach1977_kondo_lattice_weak_antiferromagnetism}. Not surprisingly,
the R2B model is unable to provide a proper description of the system
beyond the transition in the Mott insulator state (see Fig.~\ref{fig:dimer_transition_spectra}). We shall
come back to this point later on.

We should mention here that, within the IPT approximation and our numerical
precision, \(\eta\) does not seem to vanish completely at intermediate values
of \(t_\perp\) (\(\approx 0.5\)) as clearly as it does for smaller and larger
values (see Fig.~\ref{fig:transition_orderparam}). We have also used CT-QMC
at the lowest possible temperatures but the results were inconclusive due to
the very small energy scale being in competition with the low
temperature. This issue might be eventually fully resolved by better adapted methods
such as NRG-DMFT
\cite{bulla1998_numerical_renormalization_group_calculations_self} or
DMRG-DMFT \cite{garcia2004_dynamical_mean_field_theory_with}.

\subsection{Insulator to metal transition across \(U_{c_1}\)}
\label{sec:orge8aebc9}

\begin{figure}[htbp]
\centering
\includegraphics[width=\linewidth]{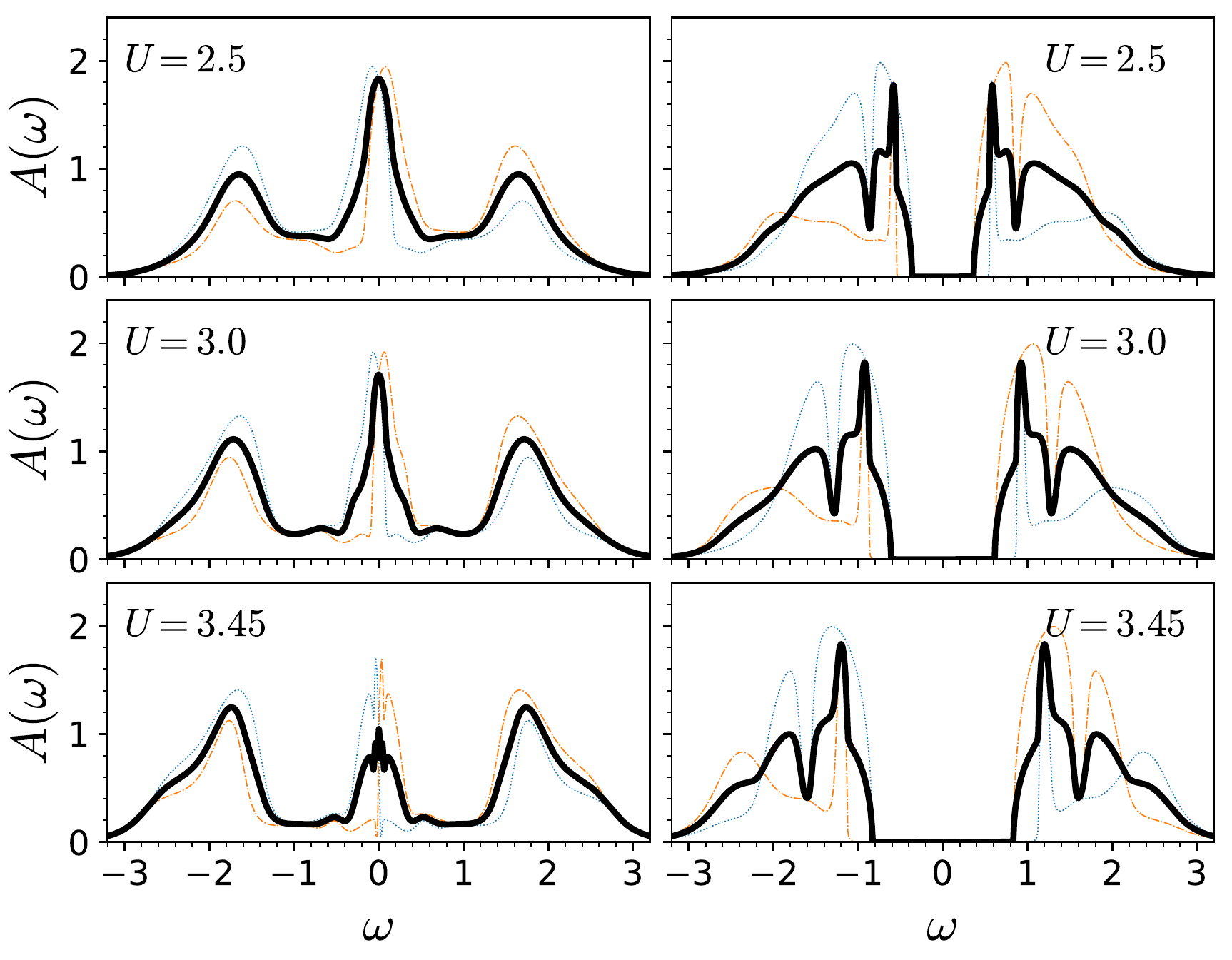}
\caption{DOS for increasing values of \(U\) and \(t_\perp=0.3\) within the coexistence region. Metallic solutions are shown on left panels and insulating ones on the right. Black lines represent the local(\(-\Im m G_{11}\)) Spectral function, which is the average of the bonding (dashed blue) and anti-bonding (dot dashed orange) bands, calculated at \(T=0\) with IPT. \label{fig:DOS_coex}}
\end{figure}

An interesting feature of the solution of the DHM is the existence of a
first order transition driven by temperature \cite{najera2017_resolving_vo2_controversy}. This
transition emerges as a consequence of two coexistent solutions found in a
region of the phase diagram of the model
\cite{moeller1999_rkky_interactions_mott_transition} as we show in Fig.
\ref{fig:DOS_coex}.

\begin{figure}[htbp]
\centering
\includegraphics[width=\linewidth]{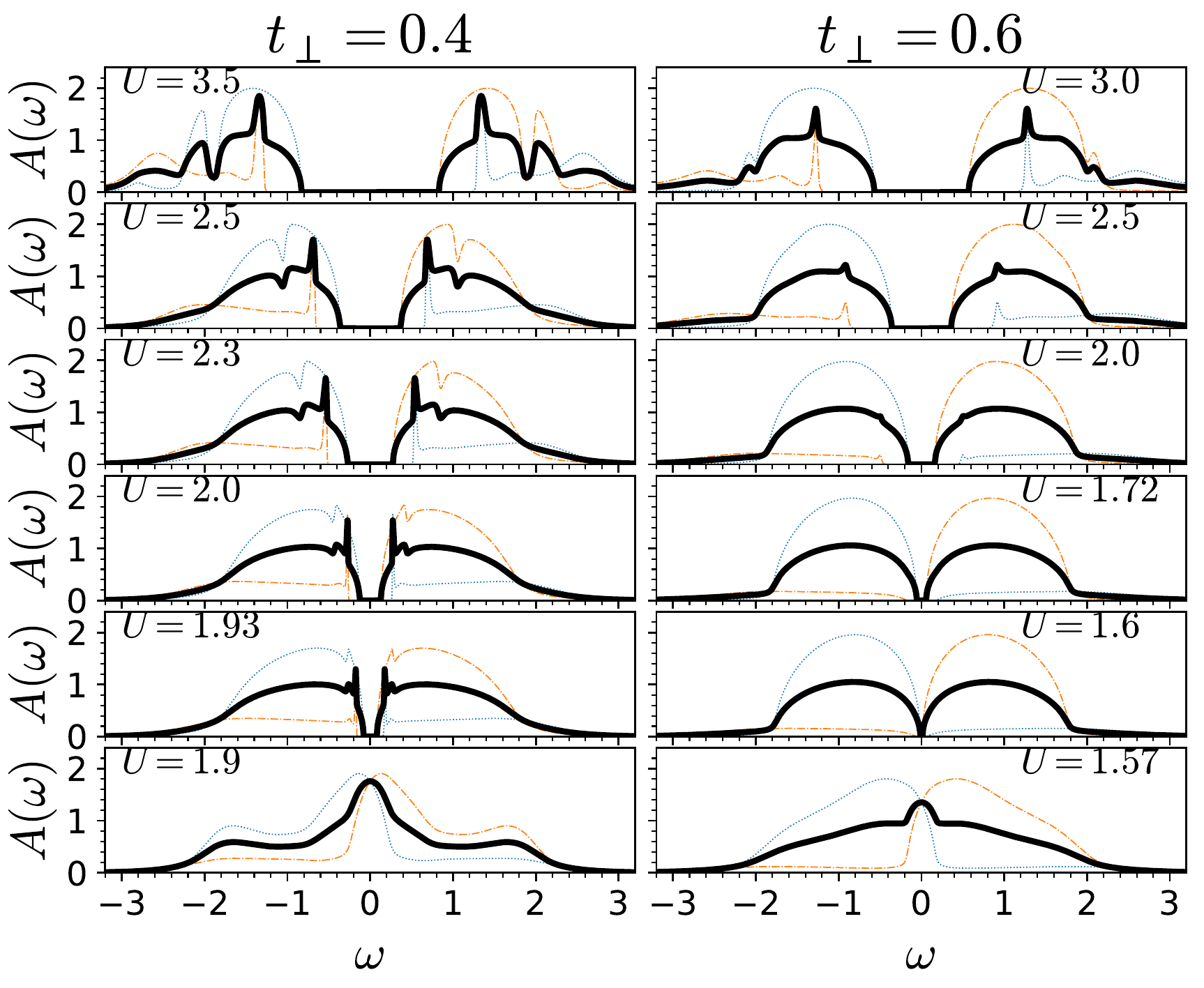}
\caption{DOS for decreasing values of \(U\) as the system crosses the \(U_{c_1}\) transition line (blue line of Fig.\ref{fig:phasediag}). Left panel is away from the tri-critical point and right panel data is close to it. Black lines represent the local(\(-\Im m G_{11}\)) Spectral function, which is the average of the bonding (dashed blue) and anti-bonding (dot dashed orange) bands, calculated at \(T=0\) with IPT. Note the discontinuous behavior as \(U_{c_1}\) is crossed. \label{fig:DOS_ins_met}}
\end{figure}

We have described above how the metallic solution collapses discontinuously
as one increases the interaction \(U\). Here we shall consider the collapse
of the insulator one as we come down from high \(U\) towards \(U_{c_1}\). The
systematic behavior of the DOS is shown in Fig.~\ref{fig:DOS_ins_met}
for two representative values of \(t_\perp\). A smaller value \(t_\perp=0.4\)
where the system crosses the \(U_{c_1}\) line and, for comparison, a larger
value \(t_\perp=0.6\) which is closer to the continuous transition line
\(U_{c_3}\).

At the lower value of \(t_\perp\) we observe that the DOS does not seem to
close the gap at the transition. Notice the choice of values of \(U\) very
close to the critical point. The
transition is clearly discontinuous, since just below \(U_{c_1}\) for \(U=1.9\)
the DOS changes abruptly, displaying a metallic state that has a large
quasiparticle peak. The line-shape of the Hubbard bands is quite peculiar
and we shall consider that feature later on. At the larger value of
\(t_\perp\) the system is still crossing the \(U_{c_1}\) line. However, and in
contrast to the previous case and within our numerical precision, the gap
seems to close continuously. Nevertheless, and different to the behavior
across the \(U_{c_3}\) line that we described before (cf
Fig.~\ref{fig:dimer_transition_spectra}), the character of the
transition from insulator to metal remains discontinuous in regard of the
DOS line-shape. Indeed, as the results show, it changes quite significantly
with only a tiny variation of \(U\) (lower two panels on the r.h.s. of
Fig.~\ref{fig:DOS_ins_met}). Also in contrast to the lower \(t_\perp\)
case, we see that the line-shape of the DOS in the insulator has
significantly less structure. This is due to the proximity of the
parameters to those of the continuous transition, therefore the first order
character becomes weaker as one approaches the \emph{tri-critical} transition
point where the \(U_{c_1}\), \(U_{c_2}\) and \(U_{c_3}\) lines meet.

\section{Mott-Peierls insulator-insulator crossover}
\label{sec:orgd233a2d}

We now turn to the central part of our study, namely, the characterization
of the multiple crossovers regimes in the DHM.

\begin{figure}[htbp]
\centering
\includegraphics[width=\linewidth]{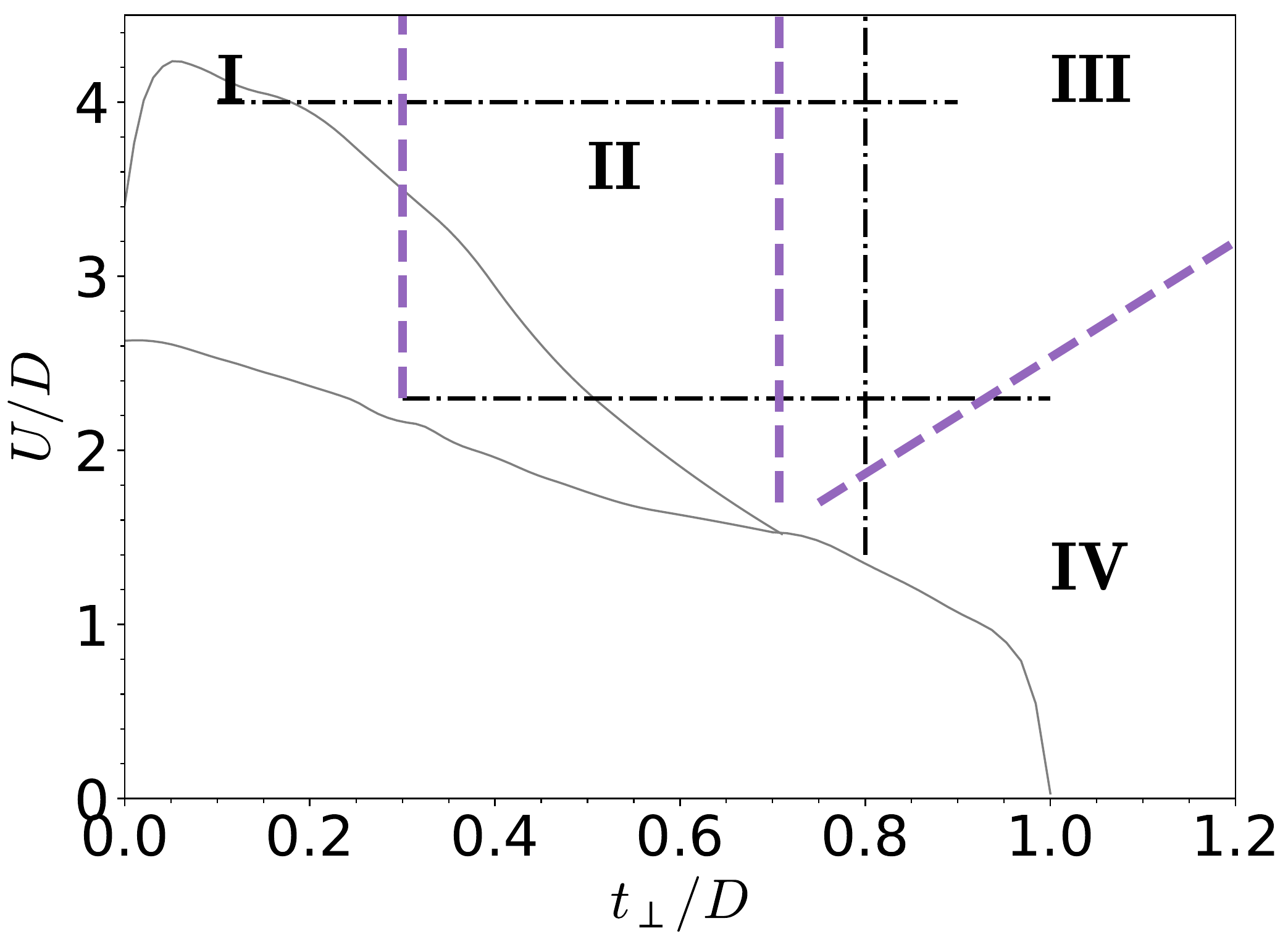}
\caption{The phase diagram of the model with the various crossover regimes \textbf{I-IV} that are described in the text. The dashed lines separating the different zones are for~reference only, since the evolution is continuous. The dash-dot lines denote the various paths across the diagram whose evolution we describe in the text. For~reference, we draw as gray lines the \(U_{c1}\), \(U_{c2}\) and \(U_{c3}\) transitions (cf Fig.~\ref{fig:phasediag}) \label{fig:crossover_diagram}}
\end{figure}

As we discussed already in the Introduction, the physical characterization
of the VO\(_2\) as Mott, Peierls or ``in-between'' has been a tricky
issue. As we shall see in this section, this can be explained by the rich
physics and subtle behavior changes that the DHM exhibits as it
crosses-over from pure Mott to pure Peierls. We may define the former with
respect to the prototypical Mott insulator that is realized in the one band
Hubbard model. As we already discussed in Section II, the DHM in the
\(t_\perp=0\) limit becomes in fact two independent copies of the
single-band Hubbard model.  In such a regime, the electrons become
localized because of the strong on-site Coulomb repulsion. This creates
``free local moments'' at every site, and the electronic structure is,
accordingly, very \emph{incoherent Hubbard bands} split by a large energy scale
\(\sim U\) \cite{georges1996_dynamical_mean_field_theory_strongly}. The other
extreme case, the pure Peierls, is identified in the DHM as the \(U=0\) limit
with the B/AB bands having a \(2t_\perp\) split, larger that the bandwidth
\(2D\). Hence, a gap in the DOS spectra opens by virtue of the
momentum-independent strong dimerization hopping amplitude. This is a pure
``band-structure'' effect as the interaction \(U\) is set to zero. In this
insulator state, the bonding and anti-bonding bands are separated and the
former is fully filled with two electrons per dimer site. The system is a
``band-insulator'', which is non-magnetic and its electronic structure
shows two parallel \emph{coherent Bloch bands}. We shall explore in this section
how the system transmutes from one regime to the other.

In Fig.~\ref{fig:crossover_diagram} we show the various regimes that the
systems exhibits as it crosses-over from the Mott to the Peierls
limit. There are four different zones, which can be well characterized. The
understanding of Zone I is key to this study. Its important feature is an
interesting \emph{thermal} crossover where spin degrees of freedom are
active. These magnetic moments are due to the Coulomb interaction and
emerge as the result of Mott localization above the \(U_{c_1}\) line at low
\(t_\perp\). Zone IV is characterized by the insulating Peierls state. As we
shall see, we can think of that state as ``orbitally polarized'' in the
B/AB basis, with correlations playing a relatively minor role. The Zones II
and III have a mix character and the evolution of the electronic structure
is quite subtle. We have therefore explored the evolution of the system
across the different zones by following the black lines that are indicated
in Fig.~\ref{fig:crossover_diagram}. We consider two parallel lines at fixed
values of \(U\) and varying \(t_\perp\). The relatively smaller \(U\) line traces
the systematic evolution from within the Mott coexistence region towards
the Peierls one. At a larger value of \(U\) we shall see that the system remains within a
Mott state even for relatively large values of \(t_\perp\). The main feature
in this case
is an interesting evolution of the electronic structure, going from
incoherent Hubbard bands (Zone I) to coherent ones (Zone III) and passing
through a \emph{mixed} state with the coexistence of coherent and incoherent
contributions (Zone II). We shall describe these various crossovers in detail in the following
subsections.

\subsection{Zone I: the singlet to free-moment crossover in the Mott state}
\label{sec:org521dcdc}

\begin{figure}[htbp]
\centering
\includegraphics[width=\linewidth]{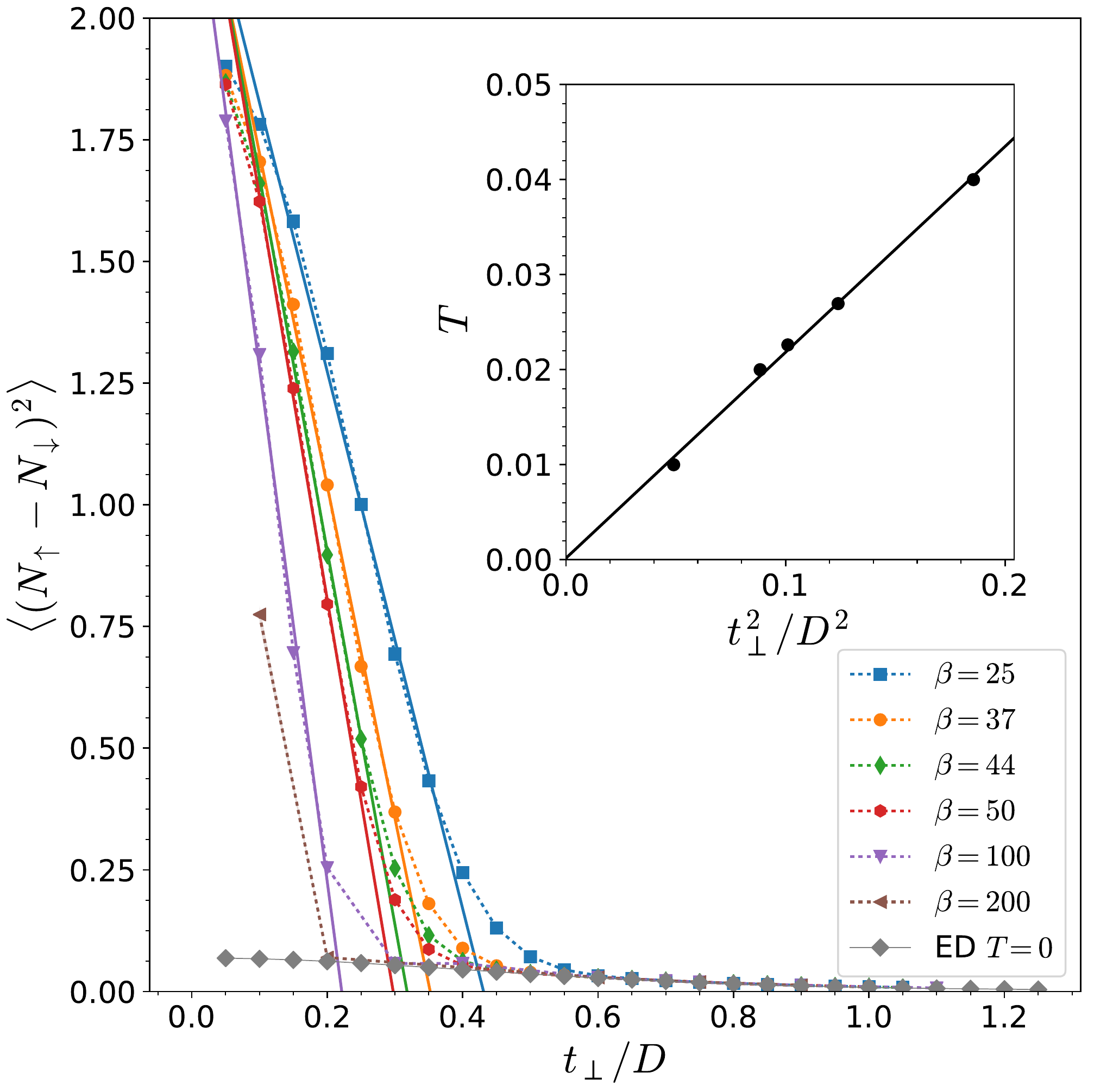}
\caption{Total magnetic moment formation as a function of \(t_\perp\) at different temperatures. Finite temperature calculation done with CTHYB, and zero temperature with ED. Inset: Temperature crossover scale for singlet pairing of the two site moments \(T^*\). This scale is proportional to the square of \(t_\perp\) consistent with its magnetic origin. \label{fig:extrapolation}}
\end{figure}

This regime at \(U>U_{c_1}\) and small \(t_\perp\) is crucial to understand the
physical behavior of the present model. The large value of the on-site Coulomb
repulsion \(U\) creates a local magnetic moment at each site of the
dimer. Then, the interaction between these moments undergoes a \emph{thermal
crossover} from a singlet ground state at \(T \to 0\) to a free-moment
regime above a low temperature scale \(T^*\). This temperature is a low
energy scale of the model, which indicates the singlet pair formation and
is two orders of magnitude smaller than the bare parameters.  In the Fig.
\ref{fig:extrapolation} we show the behavior of the total magnetic moment
formation \(\langle(N_\uparrow - N_\downarrow)^2\rangle\) =
\(\langle[(n_{1\uparrow} - n_{1\downarrow}) + (n_{2\uparrow} -
n_{2\downarrow})]^2\rangle\) as a function of \(t_\perp\) at different fixed
temperatures. At any given temperature, we observe that the moment
formation goes from a very small value at large \(t_\perp\) and suddenly has
a dramatic increase upon lowering that parameter. The reason is that the
magnetic coupling between the local moments at the two sites of the dimer
is large at bigger \(t_\perp\) so they lock into a singlet state which is
non-magnetic. When this magnetic interaction is reduced by decreasing \(t_\perp\)
the magnetic binding energy falls below the thermal energy and the singlet
state breaks down. The two local moments unbind and behave as local free
spins analogous to the Mott insulator state of the single band Hubbard
model.

\begin{figure}[htbp]
\centering
\subfloat{\includegraphics[width=0.25\textwidth]{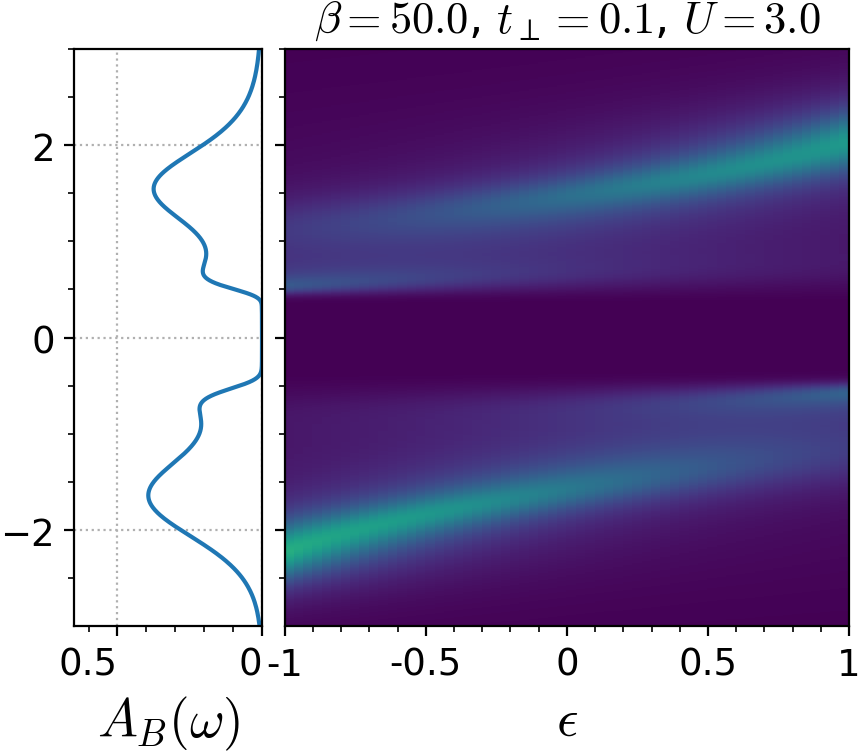}}
\subfloat{\includegraphics[width=0.25\textwidth]{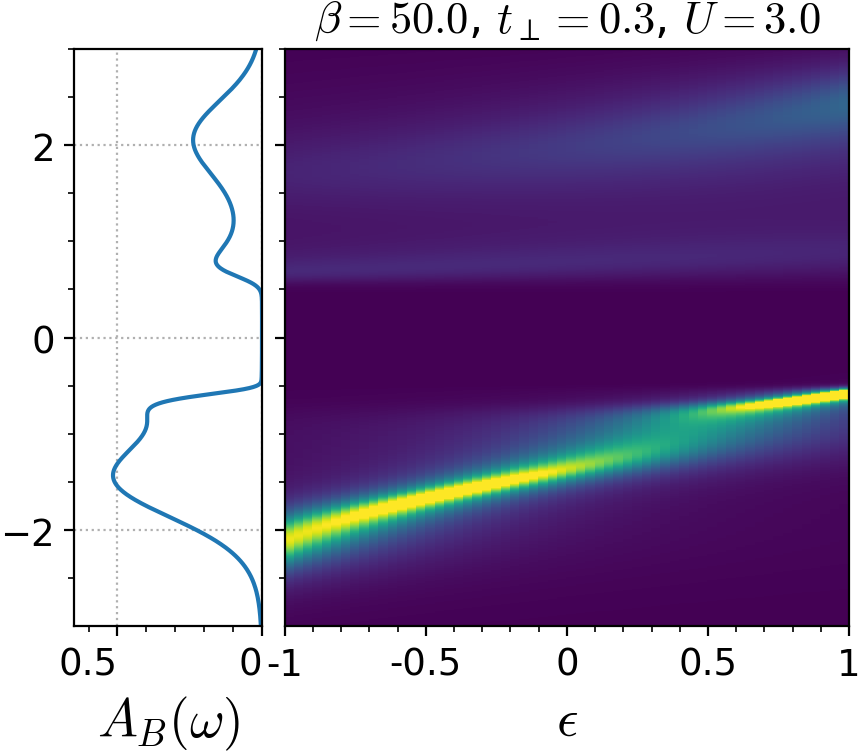}}
\caption{Intensity plot of the \emph{bonding} spectral function
$A_B(\epsilon,\omega)$ at $U$=3 and $t_\perp$=0.1 and 0.3 calculated at
$T$=0.02, using CTHYB.  In side panels we show the corresponding
DOS($\omega$), i.e. the integrated \emph{bonding} spectral function.  We
recall that in the Bethe lattice the single particle \emph{energy} plays an
analogous role as the lattice momentum $k$.  The non-interacting dispersion
is $\epsilon - t_\perp$, with $-D< \epsilon<D$.\label{fig:magnetic_DOS}}
\end{figure}

This behavior can also be clearly seen by its dramatic effect on the
electronic structure. In Fig.~\ref{fig:magnetic_DOS} we show in a color
intensity plot the \emph{bonding} spectral function dispersion
\(A_B(\epsilon,\omega)\) for the system at a fixed \(T\) and two values of
\(t_\perp\). One larger \(t_\perp=0.3\) with the two moments locked into the
singlet and a smaller one \(t_\perp=0.1\) with two unbound free-moments.

The change in the spectral function is very significant and consistent with
the magnetic state. To make the effect more explicit we focus on the
bonding spectral function. We recall that the anti-bonding is obtained by
reflection around \(\omega=0\), and the site-basis one is the average of the
two. At the higher value of \(t_\perp\) the spectra is not symmetric, we see
that the low energy band has most of the spectral weight. This signals
that the system is locked in the
singlet bound state, with the bonding band almost fully occupied. In
contrast, at lower \(t_\perp\) the state is not magnetically bound and
correspondingly we obtain a symmetric spectrum. The AB state is virtually
identical to the B one, so both have a similar occupation, which indicates
that the magnetic states are decoupled and free to fluctuate as in the Mott
insulator in the single band case. In fact, the spectral function of the
low \(t_\perp\) case that is above \(T^*\) is very similar to the incoherent
Hubbard bands of the single band Mott-Hubbard insulator
\cite{georges1996_dynamical_mean_field_theory_strongly}, which is nothing but
the \(t_\perp=0\) case.

This physical insight is a key reference to guide the discussion of the
various \(t_\perp\) and \(U\) dependent crossovers that we shall describe
next. In fact, we shall see that the emerging magnetic moments,
characteristic of the Mott localization phenomenon will show up in
different contributions to the electronic structure.

\subsection{Zone IV-III and IV-I crossovers: building correlations on the Peierls state}
\label{sec:org473af62}

\begin{figure}[htbp]
\centering
\includegraphics[width=\linewidth]{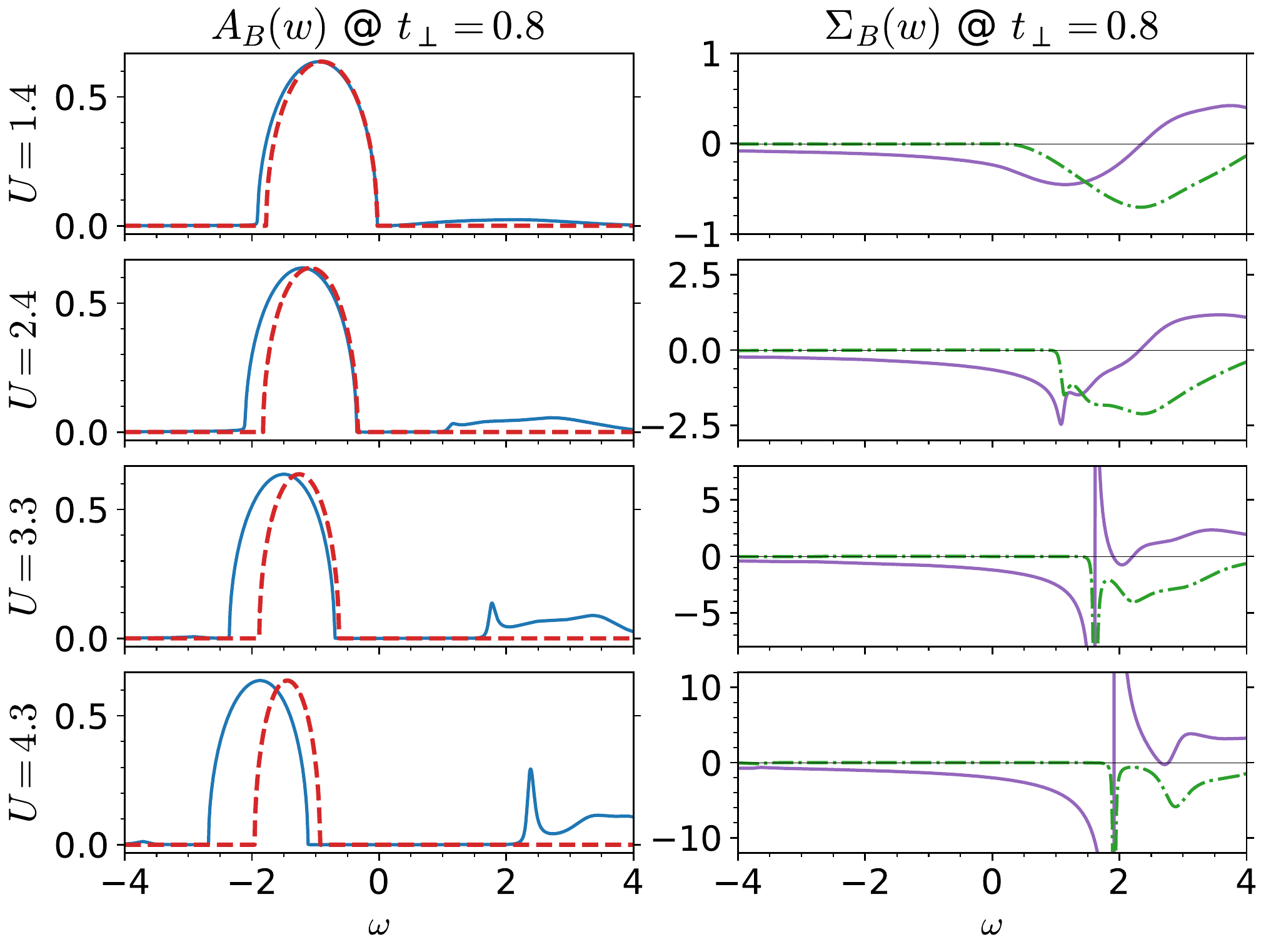}
\caption{Evolution of the bonding DOS(\(\omega\)) and corresponding change in the bonding Self-Energy at fixed \(t_\perp=0.8\)(real part in solid lines and imaginary part in dot-dashed lines) for increasing \(U\). The location of this crossover path is indicated by one of the black dash-dotted lines in Fig. ~\ref{fig:crossover_diagram}.  Red dashed lines correspond to the R2B model parametrization, which fails at large \(U\) \label{fig:IPT_EF_Uevol_ins}}
\end{figure}

We have just discussed how the local moment degrees of freedom that are
present in Zone I due to Mott localization may bind or unbind. Now we will
discuss how in the Peierls non-magnetic state (large \(t_\perp\) and \(U=0\)),
the magnetic moments gradually emerge as the correlations are increased.
To illustrate that, we plot in Fig.~\ref{fig:IPT_EF_Uevol_ins} the evolution
of the bonding orbital DOS, i.e. the \(A_B(\omega)\) spectral function for
increasing \(U\), along with its corresponding self-energy
\(\Sigma_B(\omega)\). At weak correlations, for \(t_\perp=0.8\) and \(U=1.4\),
the system is still within the Peierls insulating state in Zone IV. We
observe that the R2B model parametrization (red dashed line) provides a
rather good description. The occupation of the B state is almost complete,
so we may consider this state as fully orbitally polarized in the B/AB
basis. Accordingly, the self-energy remains smooth and small.

Increasing \(U\) the system crosses-over from Zone IV to III and we
observe qualitative changes in both, the DOS and \(\Sigma\). For \(U=2.4\)
we already see an incipiently structured spectral weight developing at
\(\omega>0\). Accordingly, the self-energy begins to develop a rapidly
varying wiggle. These developing structures become apparent at a higher
interaction strength \(U=3.3\). We clearly observe the emergence of a
quasiparticle-like resonance in the DOS, with a concomitant pole in the
self-energy. This signals the onset of a well-defined excitation and
the narrowness of the peak indicates that is essentially a localized
state. This excitation is in fact due to local moments building up at each
of the dimer sites \cite{najera2017_resolving_vo2_controversy}. Unlike in the
Mott state at low \(t_\perp\), where the local moments produce a broad
incoherent contribution to the spectral function, here the moments are
strongly coupled by the large \(t_\perp\). Therefore, they remain
Mott-localized \emph{within} the dimer but establish a local coherent (singlet)
state. Upon further increasing the interaction \(U\) the resonance gains more
spectral
weight and the strength of the pole also grows. The ``Mottness'' character
of the state increases as we see that the simple renormalized two band
parametrization fully breaks down. Notice, however, that in contrast to the
pure Mott insulator with incoherent Hubbard bands, here the lower Hubbard
band in \(A_B(\omega)\) remains fully coherent as in the Peierls insulator
case. This is evident from the imaginary part of the self-energy (bottom
right panel of Fig.~\ref{fig:IPT_EF_tpevol_inslU}), which is negligible on
the full \(\omega<0\) frequency range. We thus begin to observe the
coexistence of incoherent and coherent features in the electronic
structure, which are respectively connected to Mott and Peierls physics.

\begin{figure}[htbp]
\centering
\includegraphics[width=\linewidth]{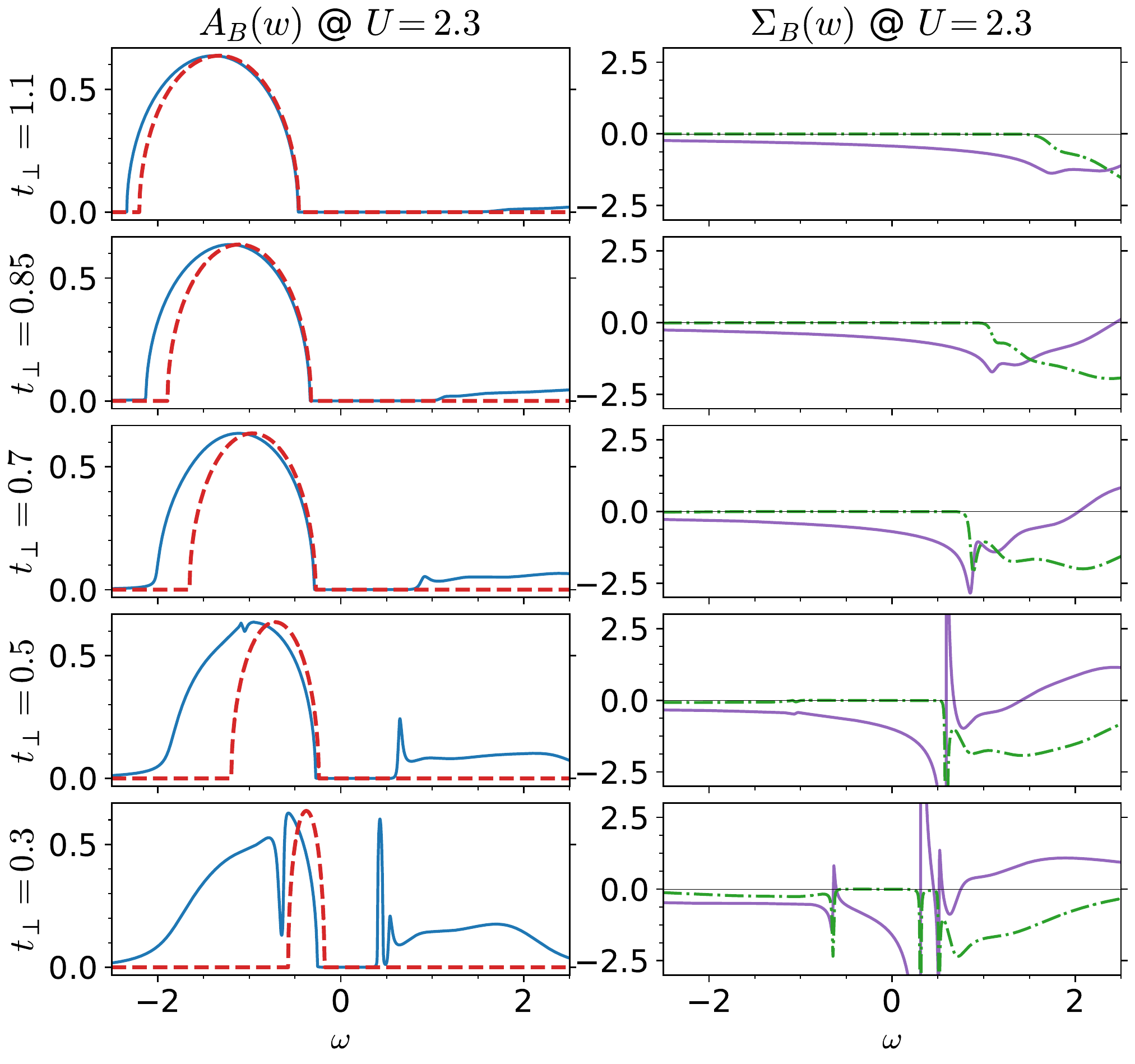}
\caption{Evolution of the bonding DOS(\(\omega\)) and corresponding change in the bonding Self-Energy(real part in solid lines and imaginary part in dot-dashed) at fixed \(U=2.3\) for decreasing \(t_\perp\). The location of this crossover path is indicated by one of the black dash-dotted lines in Fig.~\ref{fig:crossover_diagram}. Red dashed lines correspond to the R2B model parametrization, which fails at lowest \(t_\perp\). While there is a sharp narrow peak at the w>0 of the spectra (lower panel), we are not fully confident in the presence of the small secondary peak being a separate excitation. \label{fig:IPT_tpevol_ins}}
\end{figure}

We now turn to the crossover behavior from zone IV to zone I (cf
Fig.~\ref{fig:crossover_diagram}). The systematic behavior is shown for
\(A_B(\omega)\) and \(\Sigma_B(\omega)\) in Fig.~\ref{fig:IPT_tpevol_ins}.

We observe that all the features that we described before in the zone IV to
III crossover as a function of \(U\) are also present here as the system
evolves as a function of the model parameter \(t_\perp\). It crosses-over
from the Peierls insulator in zone IV towards the Mott state
in zone I. Similarly as before, we observe the emergence of a narrow
resonance in the \(\omega>0\) part of the spectra. However, a difference with
the previous crossover is that the evolution now ends close to the pure
Mott state and we see that the \(\Sigma_B(\omega)\) is non negligible at
\(\omega<0\).
One may notice that this second crossover path traverses the zone II. The
clear characterization of that regime requires the discussion of the
spectral function \(A(\epsilon,\omega)\), which we shall consider in the next
subsection.

\section{The evolution of the dimer Mott insulator: coexistent incoherent and coherent contributions}
\label{sec:org921b04e}

\begin{figure}[htbp]
\centering
\includegraphics[width=\linewidth]{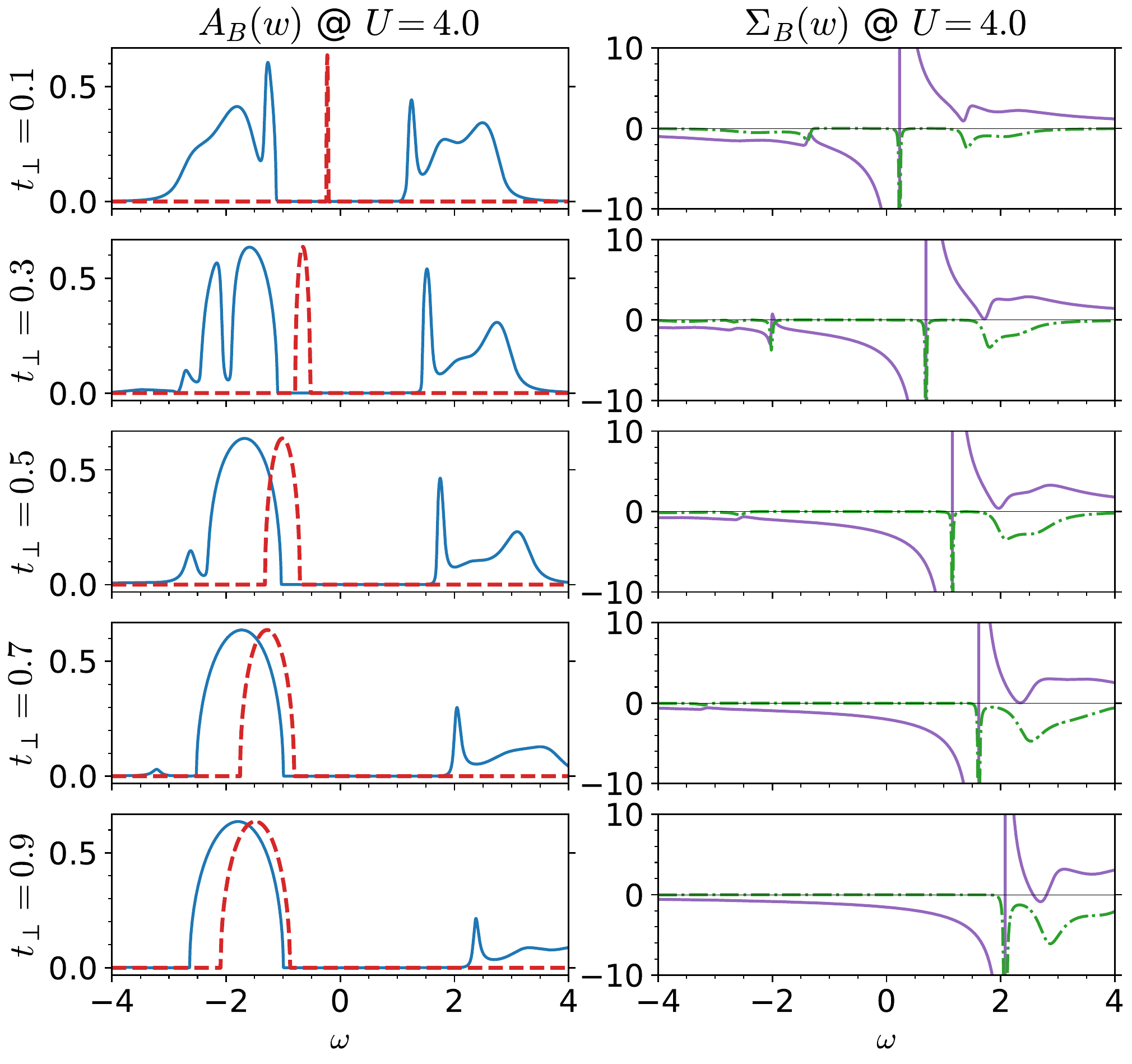}
\caption{Evolution of the bonding DOS(\(\omega\)) and corresponding change in the bonding Self-Energy(real part in solid lines and imaginary part in dot-dashed) at fixed \(U=4\) for increasing \(t_\perp\). The location of this crossover path is indicated by one of the black dash-dotted lines in Fig. \ref{fig:crossover_diagram}.  Red dashed lines correspond to the R2B model parametrization, which is always poor in this case. \label{fig:IPT_EF_tpevol_inslU}}
\end{figure}

We now finally consider the strongly correlated regime set by a relatively
large value of the interaction \(U=4\). We shall discuss the systematic
changes of the insulator state as it evolves as a function of increasing
intra-dimer hopping \(t_\perp\). As we did before, we begin considering the
bonding DOS \(A_B(\omega)\) and the corresponding self-energy
\(\Sigma_B(\omega)\), which unveils details of its mathematical structure.
The data are shown in Fig.~\ref{fig:IPT_EF_tpevol_inslU}. We recall that the
same quantities on the AB orbital are obtained by reflection around
\(\omega=0\), and that the total DOS corresponds to the average of the B and
AB. The main feature is that there is always a large gap with two main
contributions at \(\omega\sim\pm U/2\). Thus for all \(t_\perp\) we have a
large insulating gap controlled by \(U\), which is an indication of Mott
physics having a dominant role. We also see, consistent with this
observation and with our discussion in previous sections, that the unoccupied
part of \(A_B(\omega)\) always has a sharp resonance that we associated to
emergent magnetic moments. Moreover, in \(\Sigma_B(\omega)\) we always
observe the presence of a strong pole. Interestingly, we see that the
position of the pole is almost at the center of the gap at lower
\(t_\perp\). In fact, it must reach \(\omega=0\) in the limit of \(t_\perp \to 0\)
as system the becomes two independent copies of a single-band Mott insulator
\cite{georges1996_dynamical_mean_field_theory_strongly}. Thus, this strong
pole is a hallmark of the opening of a Mott gap. As we increase \(t_\perp\)
we see that the pole remains strong but evolves towards the upper edge of
the gap. This has the effect of strongly affecting the \(\omega>0\) part of
the spectrum while we observe that the \(\omega<0\), in contrast, evolves
towards the semi-circular density of states. This apparent weakening of
correlations in the lower Hubbard band can be also understood by the fact
that this band is further filled up, hence effectively moving away from the
half-filled situation. However, it would be a mistake to simply consider
this a weakly correlated state since, as we already emphasized, the gap is
large and set by the Coulomb interaction \(U\). In fact, we observe that the
R2B model parametrization (red dashes line in the
Fig.~\ref{fig:IPT_EF_tpevol_inslU}) is poor in all cases.

Another interesting feature revealed by the \(\Sigma_B(\omega)\) is that the
imaginary part, which is related to the inverse life-time is always
relatively large on the \(\omega>0\) side of the spectra while is much
smaller, or even negligible for larger \(t_\perp\) on the \(\omega<0\)
side. This indicates that the positive frequency excitations have
incoherent character (save for the sharp resonance state that we discussed
in previous sections) while the negative frequency ones are coherent. One
additional interesting feature that we would like to point out is the
complex evolution of the line shape of \(A_B(\omega)\) at small \(t_\perp\),
where \(\Im\Sigma_B\) is still non-negligible. This regime corresponds to the
crossover zone II (cf. Fig.~\ref{fig:crossover_diagram}). In order to gain
further insight into these issues we shall consider the ``momentum''
resolved spectral function \(A(\epsilon,\omega)\) along with the local DOS,
\(A(\omega)=\int d\epsilon A(\epsilon,\omega)\).

\begin{figure*}\centering
\subfloat{\includegraphics[height=0.23\textwidth]{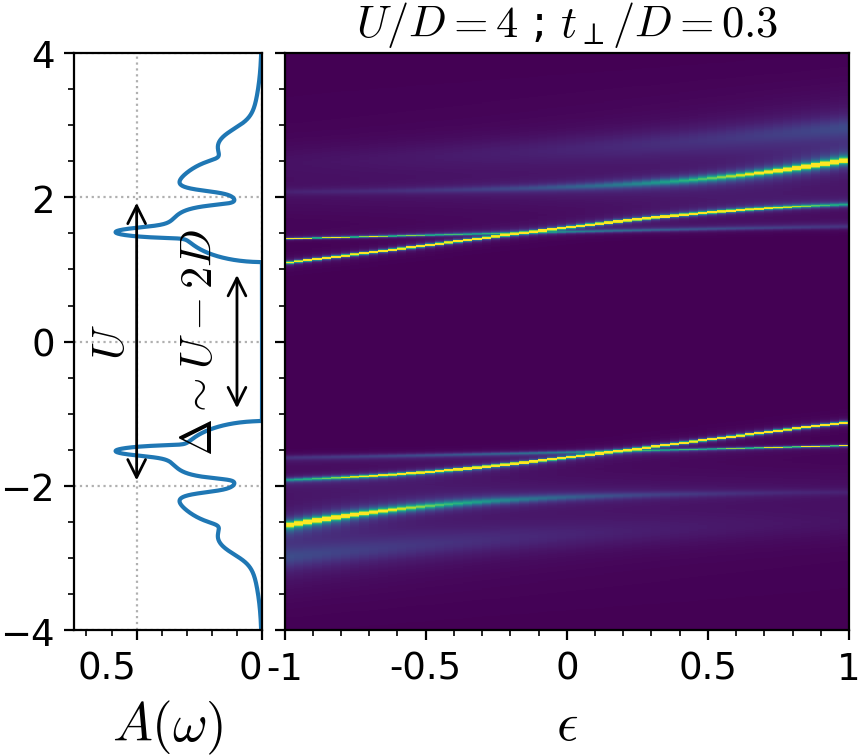}}\hfill
\subfloat{\includegraphics[height=0.23\textwidth]{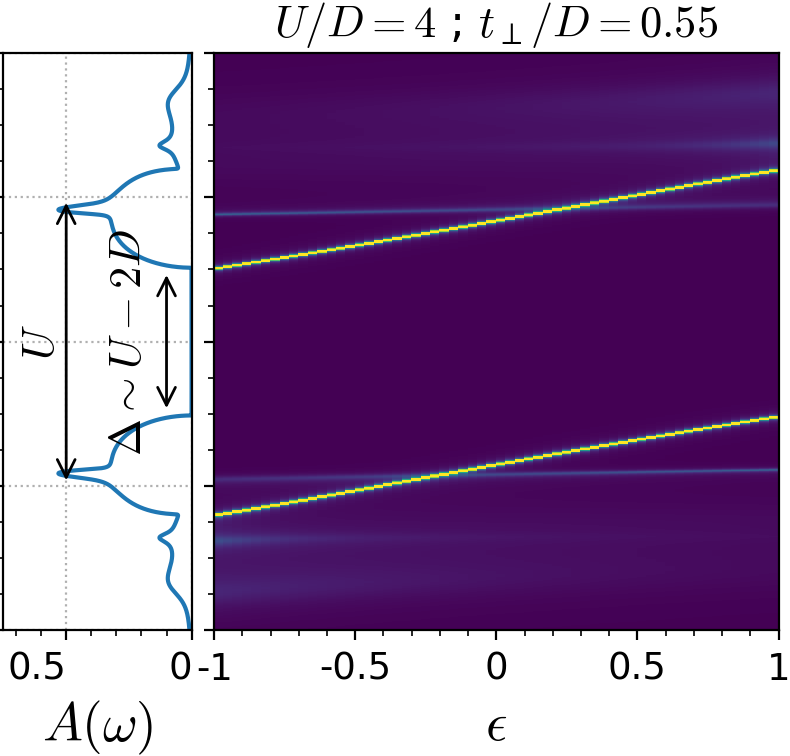}}\hfill
\subfloat{\includegraphics[height=0.23\textwidth]{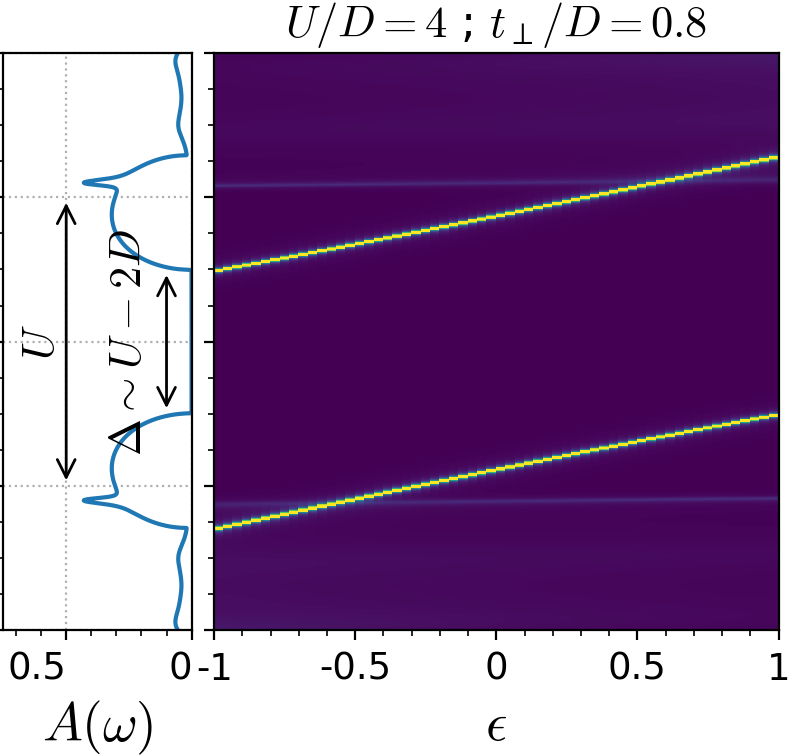}}\hfill
\subfloat{\includegraphics[height=0.23\textwidth]{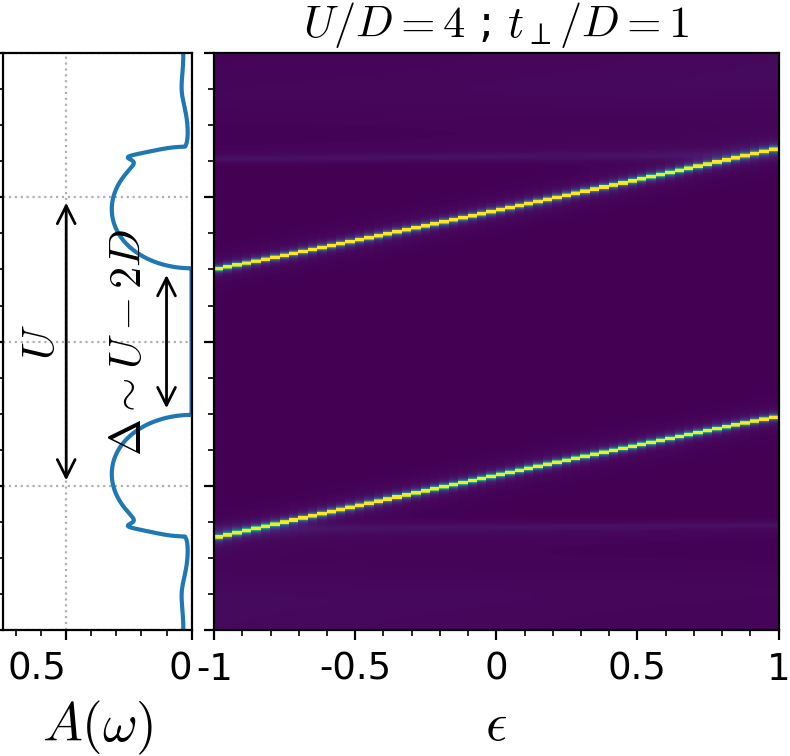}}
\caption{Local density of states $A(\omega)$ (vertical panels) and
intensity plots of the spectral function $A(\epsilon,\omega)$ of the Mott
insulating state at large $U=4$ and increasing $t_\perp$. Obtained by IPT
at $T=0$. \label{fig:spectral_dispersionTot}}
\end{figure*}

\begin{figure*}\centering
\subfloat{\includegraphics[height=0.23\textwidth]{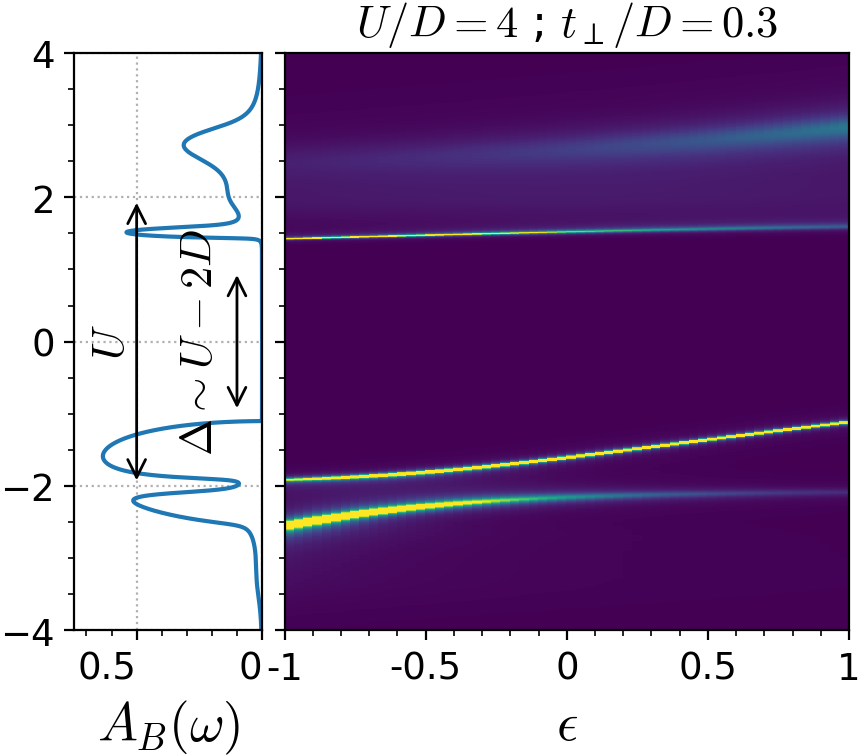}}\hfill
\subfloat{\includegraphics[height=0.23\textwidth]{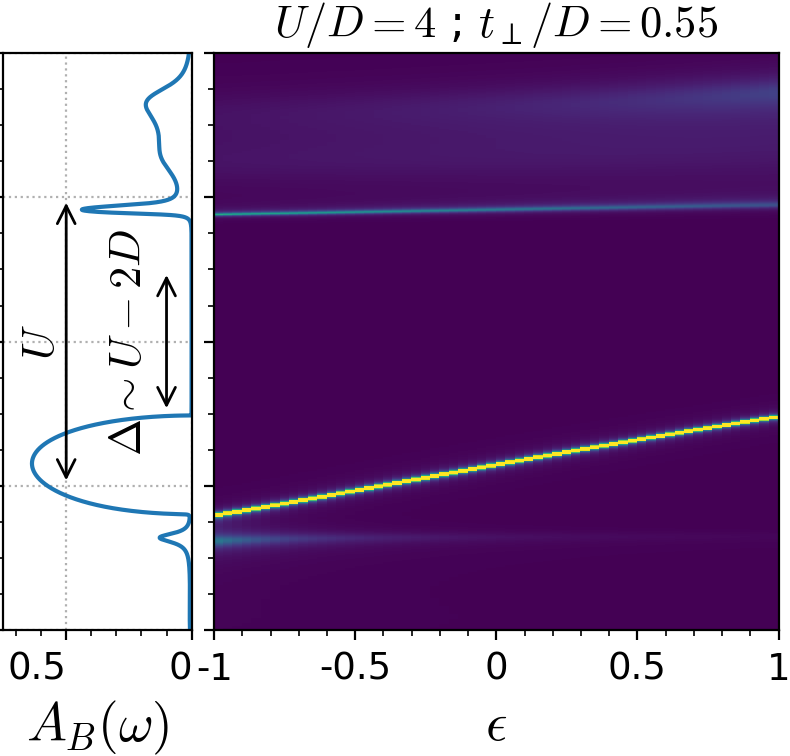}}\hfill
\subfloat{\includegraphics[height=0.23\textwidth]{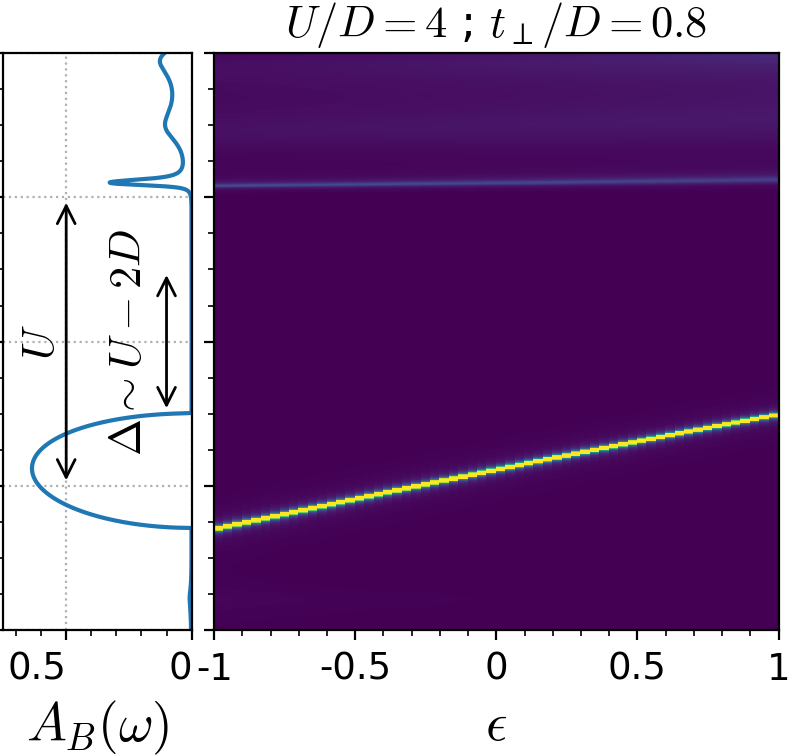}}\hfill
\subfloat{\includegraphics[height=0.23\textwidth]{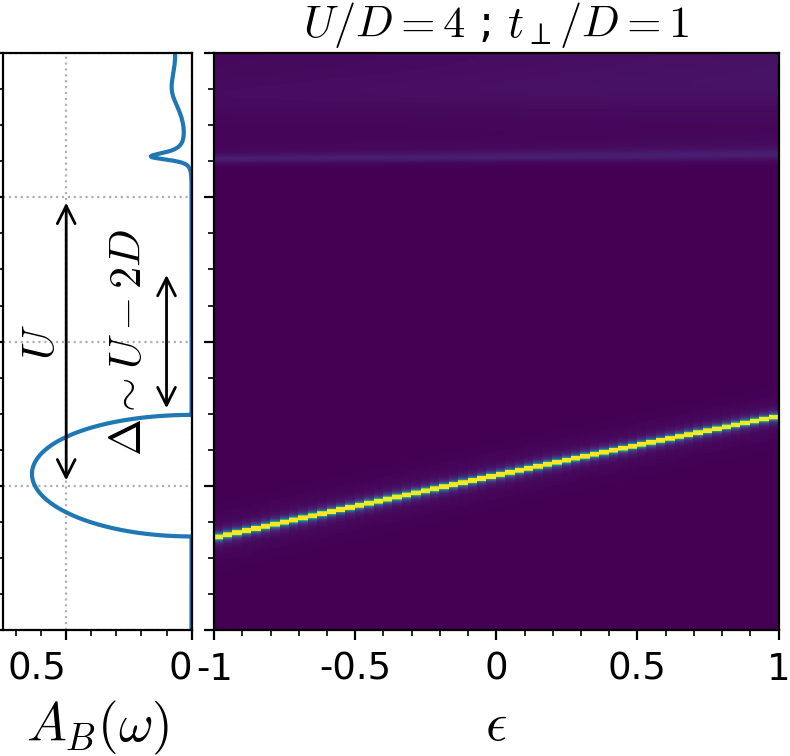}}
\caption{Bonding density of states $A_{B}(\omega)$ (vertical panels) and
intensity plots of the bonding spectral function $A_{B}(\epsilon,\omega)$
of the Mott insulating state at large $U=4$ and increasing
$t_\perp$. Obtained by IPT at $T=0$. Note these is the same data as in
Fig.\ref{fig:spectral_dispersionTot} but shown on a different basis.
\label{fig:spectral_dispersionB}}
\end{figure*}

For the sake of clarity we consider, both, the total (i.e. site basis) and
bonding orbital spectral functions. Their evolution is shown, respectively,
in Figs.~\ref{fig:spectral_dispersionTot} and
\ref{fig:spectral_dispersionB}. We should actually begin with the spectra
already shown in Fig.~\ref{fig:magnetic_DOS}, which illustrated the very low
\(t_\perp\) regime (zone I) where the system is deep in the Mott phase with
decoupled magnetic moments (at finite \(T>T^*\)). Consequently, the
\(A(\epsilon,\omega)\) displays a very incoherent electronic structure,
similar to the single band Mott state.

As we increase the \(t_\perp\) in Figs.~\ref{fig:spectral_dispersionTot} and
\ref{fig:spectral_dispersionB} we observe the systematic evolution of the
electronic structure. It always shows two roughly parallel lower and upper
Hubbard bands split by \(U\). These bands gain in coherence as \(t_\perp\) is
increased. At the end state, i.e., higher \(t_\perp\), two well-defined and
coherent contributions dominate the electronic structure. As can be seen in
the last panel of Fig.~\ref{fig:spectral_dispersionB} the main contribution
comes from the bonding state for the lower Hubbard band, and
correspondingly from the anti-bonding for the upper Hubbard one. However,
some incoherent weak intensity and weakly dispersive states can still be
observed. This state resembles the Hubbard I solution, with 2 coherent
bands, that are split by \(U\). This can be rationalized noting that in the
B/AB basis the system is strongly orbitally polarized.

More interesting are the states at lower values of \(t_\perp\). As displayed
in the first two panels we observe that the Hubbard bands develop a unique
characteristic, which is their mixed character. In fact, we observe sharper
and more coherent quasiparticle-like contributions to the electronic
structure in the inner edges of the Hubbard bands, which upon integration
lead to a peculiar line-shape for the local DOS(\(\omega\)). The outer part
of the Hubbard bands, in contrast, is strongly incoherent. The physical
interpretation of the quasiparticle states stems from the intra-dimer
magnetic coupling of the emergent moments. Within the dimer they develop a
coherent singlet state, thus remain localized and their effective mass is
heavy. The propagation of higher energy states through the lattice remains
very incoherent, as signaled by the diffuse spectral intensity which is
broad on a scale of \(\sim D\). These states with a mixed character in the
propagation of the Hubbard bands are an original feature of the DHM and
they are
absent in the single band model case. Interestingly, this regime has been
signaled as the relevant for VO\(_2\)
\cite{najera2017_resolving_vo2_controversy} and would be interesting to see
if some of its signatures may be experimentally observed in spectroscopic
studies.


\section{Conclusions}
\label{sec:org145e8b2}

We have studied in detail the solution of a basic strongly correlated
model, the dimer Hubbard model, which is relevant for VO\(_2\)
\cite{najera2017_resolving_vo2_controversy}, monoclinic transition metal
oxides MO\(_2\) \cite{hiroi2015_structural_instability_rutile_compounds} and
more generally structures with a dominating bond between a pair of
correlated metallic ions.

This model is also interesting as it is arguably the simplest realization
of a cluster DMFT problem and has a quantum impurity model that is
analogous to that of realistic DFT+DMFT calculations of
VO\(_2\). Indeed, we have seen that the solution of the DHM does
exhibit the same physical mechanism for the insulator gap opening as was
reported in DFT+DMFT studies, namely, the strong enhancement of the
intra-dimer self energy.

We provided a detailed description of the solutions in the ``coexistent
region'' where two (meta)stable states of the CDMFT equations are found,
one a metal and the other an insulator. Moreover, we described in detail
how these states break down at their respective critical lines.  We have
clarified the key role played by the intra-dimer correlation, which here
acts in addition to the familiar onsite Coulomb correlations (Mott-Hubbard)
that were already present in the one-band case. Their interplay (i.e. Kondo
screening vs RKKY) determines the physics of the metal-to-insulator
transition line as the instance where the renormalized low-energy B/AB
bands separate. This was described in terms of our R2B model parametrization,
which turned out to be always applicable in metallic side on the full \(t_\perp-U\) phase
diagram at low enough frequencies, but not in the Mott insulator state.

The simplicity of the DHM provides new and detailed physical insight and
allows us to clarify the important issue of the Mott-Peierls
crossover. This question has remained a matter of debate in DFT+CDMFT
studies for VO\(_2\). A reason may be found in the surprisingly
subtle evolution of the electronic structure with the systematic change of
model parameters. In fact, the crossover from the Mott to the Peierls limit
is non-trivial and we characterized a variety of physical
regimes. Interestingly, we found that in the Hubbard bands evolve from
purely incoherent (Mott) to purely coherent (Peierls) through a state
with unexpected mixed character. This feature can be understood as follows:
in the Mott limit, at low intra-dimer hopping \(t_\perp\), one has emergent
magnetic degrees of freedom that remain freely fluctuating above a rather small
spin singlet pairing temperature \(T^*\). Increasing the intra-dimer hopping
the moments bind into a spin singlet state and they acquire coherence
(i.e. a well-defined quantum state) within the dimer. However, the
excitations of such a state still lack coherence through the lattice. We
may think of this state as Mott-localized singlet-dimers. This state has a
mix character and is likely relevant for VO\(_2\). Upon further increase of
the intra-dimer hopping the bonding orbital becomes fully occupied as one
may think of \(t_\perp\) as an effective crystal field. Hence, the system
becomes orbitally polarized in the bonding/anti-bonding basis, which
renders the electronic structure coherent as quantum fluctuations are
frozen out. Nevertheless, even in this large \(t_\perp\) limit the gap
remains controlled by the interaction \(U\). Therefore, the state remains a
Mott insulator one at strong enough \(U\) (zone III on
Fig.~\ref{fig:crossover_diagram}), and although is in a B/AB polarized state,
it can be seen that a Hartree-Fock description fails.

Our work has uncovered a new paradigm of a non-magnetic Mott insulator, which
may be realized in structures with two strongly coupled correlated atoms,
with VO\(_2\) as a prototypical example. This Mott state has a surprising
coexistence of coherent and incoherent excitations. An open question is
whether this feature may be seen in spectroscopic studies, or if it may be
put in evidence by pump-probe experiments that may selectively excite
particles to the coherent or incoherent states.

Finally, the intra-dimer magnetic coupling provides binding of the two
electrons into a singlet state below a characteristic temperature
\(T^*\). It is an interesting open question to study the fate of such a
state upon doping the system. This situation is rather reminiscent of the
magnetic pseudo-gap state discussed in the context of the doped Mott state
of curate superconductors.

\bibliographystyle{unsrt}
\bibliography{../biblio}

\appendix
\widetext

\section{Details of the two renormalized band approximation at low frequencies}
\label{r2b_approx}
In order to derive a simple description of the low frequency electronic
structure in a large region of parameter space, it is useful to expand the
mean field equations \eqref{eq:MFE} at low energies \(\omega \to 0\) once they
are in diagonal from.

\begin{equation}
 G_{B/AB}(\omega)= \int d\varepsilon\rho(\varepsilon) [ \omega -
\varepsilon \pm t_\perp -\Sigma_{B/AB}(\omega)]^{-1}
\end{equation}

\begin{figure}[htbp]
\centering
\includegraphics[width=0.9\linewidth]{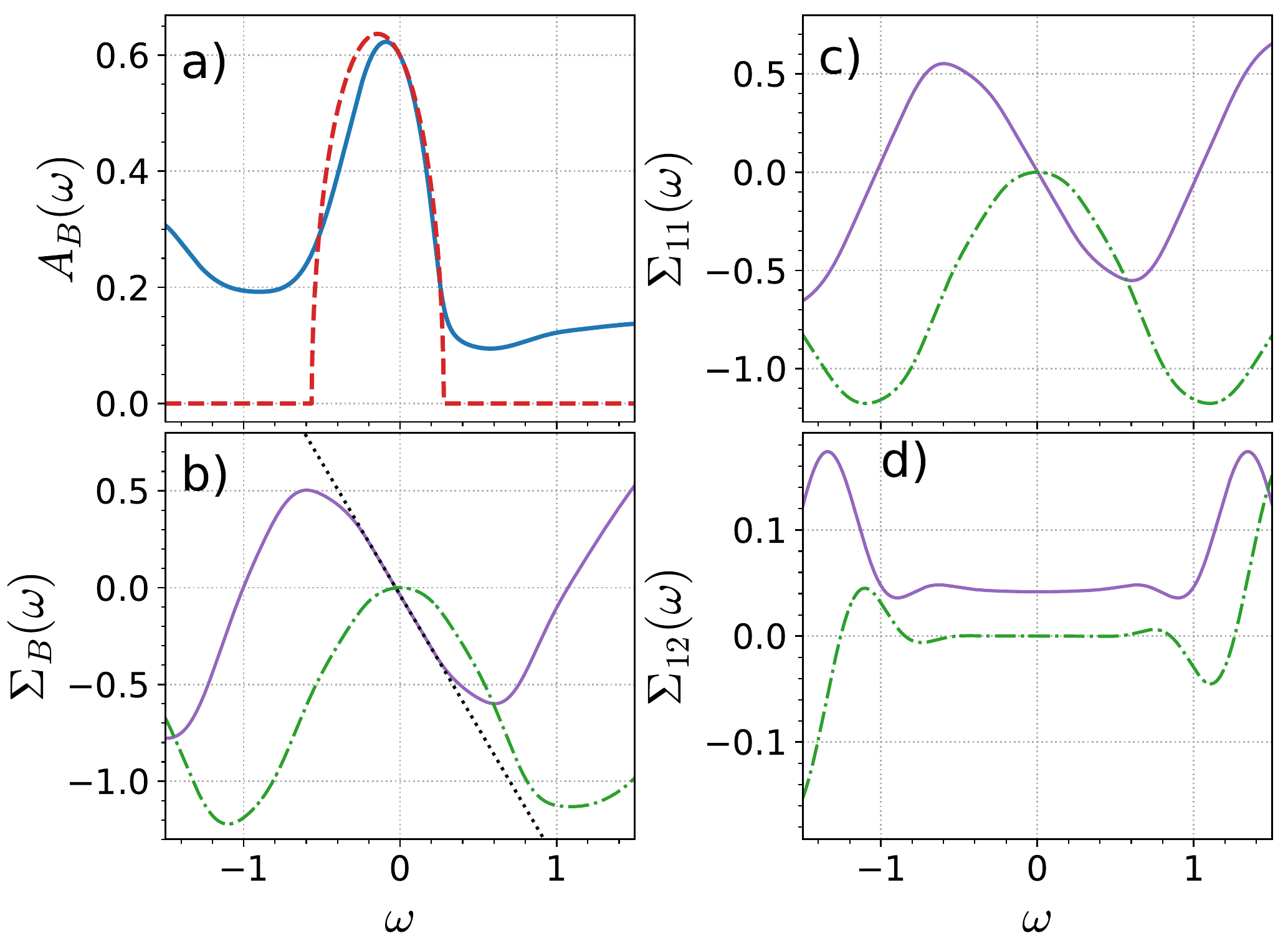}
\caption{Low energy behavior of the spectral function and Self-energy for the dimer system in the metallic phase at \(t_\perp/D=0.3\) \(U/D=2\). (a) Bonding spectral function \(A_B(\omega)\) in orange and bonding renormalized band in dashed red lines providing a good agreement with the band edge of the spectral function. (b) The corresponding self-energy in the bonding basis \(\Sigma_B\). In the site basis, (c) the same site \(\Sigma_{11}\) and (d) inter-site \(\Sigma_{12}\) Self-energies. Solid lines are the real parts and in dot dashed lines the imaginary parts. As expected in the quasiparticle regime \(\Im \Sigma\approx 0\) and the real part is linear. \label{fig:low_energy_expansion}}
\end{figure}

From this equation one approximates the Self-energy assuming it has a
well-behaved Taylor expansion up to linear order in the interval \(0 <
\omega < \omega^*\), i.e., \(\Re \Sigma'_{B/AB} (0) \approx \Re
\Sigma'_{B/AB} (\omega^*)\) we can write.

\begin{equation}
\label{eq:low_w_sigma}
\Sigma_{B/AB}(\omega)\approx\Sigma_{B/AB}(\omega=0) + \frac{\partial \Sigma_{B/AB} (\omega)}{\partial \omega} \bigg |_{w=0} \omega
\end{equation}

and call this the quasiparticle interval where this approximation holds.
The quasiparticles are long-lived provided \(\Im \Sigma_{B/AB}(\omega)
\approx 0\). In such way, we may then represent the low energy Green's
function as:

\begin{equation}
\label{eq:qp-w0}
G^{R2B}_{B/AB}(\omega)= \int d\varepsilon\rho(\varepsilon) \left[ \omega \bigg( 1-\frac{\partial\Re \Sigma_{B/AB} (\omega)}{\partial \omega} \bigg |_{0}\bigg)-
\varepsilon \pm t_\perp - \Re\Sigma_{B/AB}(0)\right]^{-1}
\end{equation}

defining as in the main text the quasiparticle residue \(Z\) by:
\begin{equation}
\label{eq:Z_appendix}
Z^{-1} \equiv 1-\frac{\partial\Re \Sigma_{B/AB} (\omega)}{\partial \omega} \bigg |_{0}=1-\frac{\partial\Re \Sigma_{11} (\omega)}{\partial \omega} \bigg |_{0}
\end{equation}
and the renormalized intra-dimer hopping:
\begin{equation}
\label{eq:tp} \tilde{t}_\perp \equiv t_\perp \mp \Re \Sigma_{B/AB}(0) =
t_\perp + \Re \Sigma_{12}(0)
\end{equation}
where the last equivalences are due to \(\Re\Sigma_{11}(0)=0\) and
\(\frac{\partial\Re \Sigma_{12} (\omega)}{\partial \omega} \bigg |_{0}=0\)
as can be verified in figure~\ref{fig:low_energy_expansion}. Thus, we obtain
a renormalized two-band (R2B) representation of the electronic structure at
low \(\omega\) in terms of two quasiparticle bands. Their corresponding DOS
is composed of two narrowed semicircles of width \(2\tilde{D} = 2ZD\) and
split by \(2Z\tilde{t}_\perp\),

\begin{equation}
\label{eq:low_energy_dos_appendix}
\rho^{R2B}_{B/AB}(\omega)=\, \frac{2}{\pi D^2}\,
\sqrt{D^2-(\frac{\omega}{Z} \pm \tilde{t}_{\perp})^2}
\end{equation}

Under this approximation the total spectral weight of this renormalized
bands is not one any more but \(Z<1\) and the rest of the spectral function
outside the quasiparticle regime has totally vanished. However, this
renormalized two band system is well capable of representing the low energy
states of the system especially the band edges of the quasiparticle peaks
through which we are able to quantify the metal to insulator transition, as
described in the main text. This simple low frequency description in terms
of two bands is accurate enough in a large region of the phase diagram,
where the \(\Re \Sigma_{B/AB}\) is well-behaved as mentioned above. These
regions include the weakly correlated limit but also the strongly
correlated metal.

It is important to recognize that the quasiparticle residue \(Z\) defined in
eq. \eqref{eq:Z_appendix} does not follow the standard definition of a
Landau-Fermi quasiparticle. In such case one would first find
quasiparticles by finding the poles \(\omega^*(\epsilon)\) of the spectral
function given by the solution of the equation

\begin{equation}
\label{eq:qp-band}
\omega^* - \epsilon \pm t_\perp - \Re \Sigma_{B/AB} (\omega^*) = 0
\end{equation}

In that case one finds the quasiparticle residue for each Landau
quasiparticle at every instance of the spectral dispersion, for each of the
composing bands of the system.

\begin{equation}
\label{eq:landau_Z}
Z^\epsilon_{B/AB} \equiv \left[ 1-\frac{\partial\Re \Sigma_{B/AB} (\omega)}{\partial \omega} \bigg |_{\omega=\omega^*(\epsilon)} \right]^{-1}
\end{equation}

This treatment is unnecessary for our current specific purposes since we
are working with an energy averaged spectral function and in our
particle-hole symmetric half-filled case the simple expansion around
\(\omega=0\) provides and excellent description of the low energy features of
the spectral function in particular the band edges, which we use to
quantify the transition.

\section{Proof that the IPT solution is exact at \(T=0\) for the isolated dimer (atomic limit of the DHM)}
\label{sec:org9c2f896}

At the heart of the dimer lattice problem is a Hydrogen molecule motive
(the dimer) which repeats in the lattice. Isolating this molecule in the
limit \(t\rightarrow 0\) the governing Hamiltonian is reduced to:

\begin{equation}
\label{eq:dimer_hamiltonian}
H = t_\perp \sum_\sigma (c^\dagger_{1\sigma}c_{2\sigma} + c^\dagger_{2\sigma}c_{1\sigma})
+ U (n_{1\uparrow}n_{1\downarrow}+n_{2\uparrow}n_{2\downarrow})
- \mu \sum_{\alpha,\sigma}n_{\alpha\sigma}
\end{equation}

where \(c_{\alpha\sigma}\) annihilates and electron from the dimer orbital
\(\alpha=\{1,2\}\) and with spin \(\sigma=\{\uparrow,
\downarrow\}\). \(U>0\) is the on-site repulsive Hubbard interaction and
\(t_\perp>0\) the dimer hybridization. The chemical potential is fixed to
ensure half-filling at \(\mu=U/2\).

In this work the fermionic arrangement convention is to order states by
their spin projection. Thus, the full occupation is described by the
many-body state vector:

\begin{equation}
|1\uparrow 2\uparrow 1\downarrow 2\downarrow \rangle = c^\dagger_{1\uparrow} c^\dagger_{2\uparrow} c^\dagger_{1\downarrow} c^\dagger_{2\downarrow} |\emptyset\rangle
\end{equation}

The single particle sector has only 4 states, as one has 4 particles. After
diagonalizing this block, we only find 2 distinct energy levels since up
and down spin are degenerate. We call this 2 levels the bonding(B) and
anti-bonding(AB) levels.

\begin{subequations}
\label{eq:Bonding_anti_bonding}
\begin{align}
|B\sigma\rangle &= \frac{1}{\sqrt{2}}\left ( | 1\sigma \rangle - | 2\sigma \rangle \right ) & E_{B}=-\frac{U}{2} - t_\perp \\
|AB\sigma\rangle &= \frac{1}{\sqrt{2}}\left ( | 1\sigma \rangle + | 2\sigma \rangle \right ) & E_{AB}=-\frac{U}{2} + t_\perp
\end{align}
\end{subequations}

The ground-state is in the \(N=2\) sector and it is non-degenerate \(\forall U
> 0\) and \(\forall t_\perp > 0\). Its energy eigenvalue and eigenvector
are:

\begin{subequations}
\label{eq:dimer_gs}
\begin{align}
E_{GS} &= - \frac{U}{2} - \frac{1}{2} \sqrt{U^{2} + 16 t_\perp^{2}}\\
|GS\rangle &= \frac{1}{a} \left (
\left ( |1\uparrow 2\downarrow \rangle + |2\uparrow 1\downarrow \rangle \right )
- \frac{4 t_\perp}{U + b}
\left ( |1\uparrow\downarrow \rangle + |2\uparrow \downarrow \rangle\right )\right )
\end{align}
\end{subequations}

where \(a=\sqrt{\frac{32 t_\perp^{2}}{(U + b)^{2}} + 2}\) and
\(b=\sqrt{U^{2} + 16 t_\perp^{2}}\). It becomes clear from this, that as the
local Coulomb interaction is raised, the double occupation is reduced in the
system. The zero temperature Green's function of the dimer can be obtained
through the Lehmann representation by:

\begin{equation}
\label{eq:zeroT_Lehmann}
G_{\alpha\beta\sigma}(\omega) =
\sum_m \frac{\langle GS | c_{\alpha\sigma} | m_{N=3} \rangle
 \langle m_{N=3} | c_{\beta\sigma}^\dagger | GS \rangle}{\omega - (E_m - E_{GS})} +
\sum_m \frac{\langle m_{N=1} | c_{\alpha\sigma} | GS \rangle
 \langle GS | c_{\beta\sigma}^\dagger | m_{N=1} \rangle}{\omega - (E_{GS} - E_m)}
\end{equation}

which reduces for the local (\(G_{11}\)) and inter-site (\(G_{12}\))
Green's functions into:

\begin{subequations}
\label{eq:gf_local_entries}
\begin{align}
G_{11\sigma} &= \frac{1}{a^2} \left[
 \frac{\left( 1 - \frac{4 t_\perp}{U+b}\right)^{2}}{a^{2} \left(b + 2 t_\perp + 2 \omega\right)}
- \frac{\left( 1 - \frac{4 t_\perp}{U+b}\right)^{2}}{a^{2} \left(b + 2 t_\perp - 2 \omega\right)}
+ \frac{\left( 1 + \frac{4 t_\perp}{U+b}\right)^{2}}{a^{2} \left(b - 2 t_\perp + 2 \omega\right)}
+ \frac{\left( 1 + \frac{4 t_\perp}{U+b}\right)^{2}}{a^{2} \left(- b + 2 t_\perp + 2 \omega\right)}
\right] \\
G_{12\sigma} &= \frac{1}{a^2} \left[
 \frac{\left( 1 - \frac{4 t_\perp}{U+b}\right)^{2}}{\left(b + 2 t_\perp + 2 \omega\right)}
+ \frac{\left( 1 - \frac{4 t_\perp}{U+b}\right)^{2}}{\left(b + 2 t_\perp - 2 \omega\right)}
- \frac{\left( 1 + \frac{4 t_\perp}{U+b}\right)^{2}}{\left(b - 2 t_\perp + 2 \omega\right)}
+ \frac{\left( 1 + \frac{4 t_\perp}{U+b}\right)^{2}}{\left(- b + 2 t_\perp + 2 \omega\right)}
\right]
\end{align}
\end{subequations}

The Self-Energy is obtained by solving Dyson's equation, which is a Matrix
equation of the form

\begin{equation}
\label{eq:dyson}
\left[\begin{matrix} \Sigma_{11} & \Sigma_{12}\\ \Sigma_{12} & \Sigma_{11}\end{matrix}\right]
=
\left[\begin{matrix} \omega & - t_\perp\\- t_\perp & \omega \end{matrix} \right] ^{-1}
-\left[\begin{matrix} G_{11} & G_{12}\\ G_{12} & G_{11}\end{matrix}\right] ^{-1}
\end{equation}
and results in:

\begin{subequations}
\label{eq:dimerT0-self-energy}
\begin{align}
\Sigma_{11}&= \frac{U^2}{4}\frac{\omega}{\omega - 9 t_\perp^2} =
\frac{U^2}{8}\left(\frac{1}{\omega+3t_\perp} + \frac{1}{\omega-3t_\perp}\right)\\
\Sigma_{12}&= \frac{U^2}{4}\frac{3t_\perp}{9 t^2_\perp - \omega^2} =
\frac{U^2}{8}\left(\frac{1}{\omega+3t_\perp} - \frac{1}{\omega-3t_\perp}\right)
\end{align}
\end{subequations}

The IPT scheme is drastically simplified in the isolated molecule case as
it becomes a single iteration procedure. The Self-Energy is directly
approximated by the second order diagram:

\begin{equation}
\label{eq:Sigma_tau}
\Sigma_{\alpha\beta}(i\omega_n) \approx - U^2 \int_0^\beta
G^0_{\alpha\beta}(\tau)G_{\alpha\beta}^0(-\tau)G_{\alpha\beta}^0(\tau) e^{i\omega_n \tau} d\tau
\end{equation}

In this case one does not include the Hartree term as for the particle-hole
symmetric case it is exactly canceled by the chemical potential. The IPT
equation \eqref{eq:Sigma_tau} can be conveniently~reformulated into real
frequencies by the analytical continuation and we can focus only on the
imaginary part only needing the spectral functions \cite{rozenberg1994_mott_hubbard_transition_infinite_dimensions}:

\begin{equation}
\label{eq:IPT_Sigma_withP}%
\Im m \Sigma_{\alpha\beta} (\nu) = -\pi U^2 \int dw dw' [A_{\alpha\beta}^-(w) A_{\alpha\beta}^+(w') A_{\alpha\beta}^-(\nu-w+w')+A_{\alpha\beta}^+(w) A_{\alpha\beta}^-(w')A_{\alpha\beta}^+(\nu-w+w')]
\end{equation}

Where
\begin{subequations}
\label{eq:Ap_Am}%
\begin{alignat}{3}
A_{\alpha\beta}^+(w) &= \theta(w)A_{\alpha\beta}(w) = -\theta (w) \Im m G_{\alpha\beta}^0(w) / \pi &&= \frac{1}{2} \left(\delta(w-t_\perp) + \delta(w+t_\perp)\right) \\
A_{\alpha\beta}^-(w) &= \theta(-w)A_{\alpha\beta}(w) = -\theta (-w) \Im m G_{\alpha\beta}^0(w) / \pi &&= \frac{1}{2} \left(\delta(w-t_\perp) - \delta(w+t_\perp)\right)
\end{alignat}
\end{subequations}

here \(\theta(w)\) is the Heaviside step function. Replacing into equation
\eqref{eq:IPT_Sigma_withP}

\begin{align*}
\label{eq:IPT_Sigma_11}
\Im m \Sigma_{11} (\nu) = -\pi \frac{U^2}{4} \int dw dw' [
 &\theta(-w)\delta(w+t_\perp) \theta(w') \delta(w'-t_\perp) A^-(\nu-w+w') + \\
 &\theta(w) \delta(w-t_\perp)\theta(-w') \delta(w'+t_\perp) A^+(\nu-w+w')] \\
= -\pi \frac{U^2}{4} \int dw' [
 &\theta(w') \delta(w'-t_\perp) A^-(\nu+t_\perp+w') + \\
 &\theta(-w') \delta(w'+t_\perp) A^+(\nu-t_\perp+w')] \\
= -\pi \frac{U^2}{4} [&A^-(\nu+2t_\perp) + A^+(\nu-2t_\perp)]\\
= -\pi \frac{U^2}{8} [&\theta(-(\nu+2t_\perp))(\delta(\nu -t_\perp)+\delta(\nu +3t_\perp)) + \\
 & \theta(\nu-2t_\perp)(\delta(\nu - 3t_\perp) + \delta(\nu - 3t_\perp))] \\
\end{align*}

\begin{equation}
\label{eq:Sigma_11}
\Im m \Sigma_{11} (\nu) = -\pi \frac{U^2}{8} [\delta(\nu +3t_\perp) + \delta(\nu - 3t_\perp)]
\end{equation}

The real part can be obtained by the Kramers-Kronig relation. A similar
procedure is followed to find \(\Sigma_{12}\), after which one recovers the
exact expressions as presented in \eqref{eq:dimerT0-self-energy}. Thus, one
can assert that the IPT approximation is exact in the zero temperature and
\(t=0\) limit.
\end{document}